\documentclass[twocolumn,pra,superscriptaddress,groupedaddress]{revtex4-2} 

\usepackage[T1]{fontenc}

\usepackage{amsmath,amsfonts,bm,amssymb}
\usepackage{graphicx}  
\usepackage{color}
\usepackage{extarrows}
\usepackage{enumitem}
\usepackage{empheq}
\usepackage{tensor}
 \usepackage[scr=boondoxupr]{mathalfa} 
\usepackage{bbm}

\newcommand{\ud}{{\mathrm d}}

\newcommand{\B}{\mbox{\tiny B}}

\newcommand{\tB}{\mbox{\tiny B}}

\newcommand{\tS}{\mbox{\tiny S}}

\newcommand{\T}{\mbox{\tiny T}}

\newcommand{\SB}{\mbox{\tiny SB}}

\newcommand{\la}{\langle}
\newcommand{\ra}{\rangle}

\newcommand{\TF}{\mbox{\tiny TF}}

\newcommand{\Sec}[1]{Sec.\,\ref{#1}}

\newcommand{\nl}{\nonumber \\}
\newcommand{\be}{\begin{equation}}
\newcommand{\ee}{\end{equation}}
\newcommand{\bsube}{\begin{subequations}}
\newcommand{\esube}{\end{subequations}}
\newcommand{\Eq}[1]{Eq.\,(\ref{#1})}
\newcommand{\Eqs}[1]{Eqs.\,(\ref{#1})}

\newcommand{\RN}[1]{%
  \textup{\uppercase\expandafter{\romannumeral#1}}%
}

\allowdisplaybreaks[1]

\definecolor{darkblue}{RGB}{0, 56, 102}

\usepackage{hyperref}  
\hypersetup{hidelinks,
	colorlinks=true,
	allcolors=black,
	pdfstartview=Fit,
	breaklinks=true
}

\begin{document}

\title{An Exact Operator Formulation of Statistical Quasiparticle Theory for Nonlinear System--Bath Coupling}

\author{Yu Su}
\email{suyupilemao@mail.ustc.edu.cn}
\author{Yao Wang}
\email{wy2010@ustc.edu.cn}
\affiliation{
  Hefei National Research Center for Physical Sciences at the Microscale, University of Science and Technology of China, Hefei, Anhui 230026, China
}

\date{\today}

\begin{abstract}
We develop an operator formulation of statistical quasiparticle theory for open quantum systems with nonlinear system--bath coupling, extending beyond the linear coupling setting of the standard Gaussian influence functional formulation. For Gaussian baths, the statistical quasiparticles, termed dissipatons, are explicitly constructed as collective operators in the thermofield bath space. Our construction provides a microscopic operator basis for the dissipaton algebra and generates hierarchical equations of motion directly from the Liouville equation. Nonlinear interactions are incorporated through a recursive generalized Wick's theorem, enabling automatic generation of hierarchical equations for polynomial couplings. The resulting framework connects explicit bath operator representations to nonperturbative, non-Markovian dynamics and unifies the treatment of linear and nonlinear system--bath interactions.
\end{abstract}

\maketitle

\section{Introduction}

Quantum mechanics of open systems has been a focus of research since the early days of quantum theory and has become an essential component of modern science, particularly in chemical physics, condensed matter physics, and quantum information science \cite{Wei21,Bre02}. The core problem of open quantum dynamics is to determine the time evolution of a system ($H_{\tS}$) coupled to its environment ($H_{\B}$), where quantum decoherence, quantum dissipation, and nonequilibrium transport arise due to the system--environment interaction ($H_{\SB}$).  The system and environment together form a closed system with time evolution governed by the Schrödinger equation, or the equivalent von Neumann--Liouville equation ($\hbar \equiv 1$),
\begin{align}\label{Liouville}
  \dot\rho_{\T}(t) = -i [H_{\T},\rho_{\T}(t)] = -i [H_{\tS} + H_{\B} + H_{\tS\tB},\rho_{\T}(t)].
\end{align}
Here the system--bath interaction is generally expressed as direct product of system and bath interacting modes, namely $H_{\SB} = \sum_{a}\hat Q_a\otimes\hat F_a$. The most prevalent model of environment is the Gauss--Wick bath, denoted as \cite{Fey63118,Cal83374,Leg871} 
\begin{align*}
  H_{\B} = \sum_{j}\frac{\omega_j}{2}(\hat p_j^2 + \hat x_j^2) \quad\text{and}\quad \hat F_a = \sum_{j} c_{aj}\hat x_j.
\end{align*}
Feynman and Vernon introduce the influence functional formalism to establish a universal and nonperturbative framework for open quantum systems \cite{Fey63118,Cal83374}. In the Gauss--Wick setting, they show that the influence functional describing all the non-Markovian effects of the environment on the system only depends on the bath correlation functions, $\la\hat F_a^{\B}(t)\hat F_b^{\B}(0)\ra_{\B}$, where $\hat F_a^{\B}(t)$ is the Heisenberg operator defined with $H_{\B}$ and $\la(\cdot)\ra_{\B}$ is the average over the thermal state of the bath. Various numerical methods are proposed to solve the influence functional. There are two main widely used approaches. One is evaluating the influence functional in the path integral representation directly, such as quasi-adiabatic path-integral method  \cite{Mak954600,Mak954611,Str183322}. The other is the differential equivalence of the influence functional, the hierarchical equations of motion (HEOM) method \cite{Tan89101}, which is based on the exponential decomposition when performing derivative on the influence functional. With the development of efficient algorithms, the influence functional formalism and related methods has been applied to a wide range of problems, making great progress in the study of condensed matter physics, chemical physics, and quantum biology. Tensor representations have extended their numerical reach \cite{Shi18174102,Bor215397}, alongside complementary wave-function methods \cite{Bec001,Ren22e1614}.

However, the influence functional formalism has some theoretical limitations. On the one hand, the formalism focuses on the influence of bath on system's dynamics, which is not direct to acquire bath's dynamical information. For example, in order to obtain the system--bath correlated quantities, such as heat current and bath absorption spectra, one has to employ the nonequilibrium Green's function formalism or the system--bath entanglement theorem. On the other hand, the influence functional is only analytically solvable for the linear system--bath coupling since higher order couplings, for example $\hat Q_{ab}\hat F_a\hat F_b$, will lead to algebraic complexity when performing Wick's theorem \cite{Cas911960}. Such nonlinear couplings play an important role in modeling physical mechanism, such as the superconduct phenomena in the quadratic electron--phonon models \cite{Rag23L121109,Han24226001,Zha25134504}, vibronic spectroscopy with Duschinsky effects \cite{Zhu25234103}, spin--lattice relaxation \cite{Bi24024105}, spin shielding in the Kondo model \cite{Kon6437,Phi12}, impurity scattering in the Luttinger liquid \cite{Wu06106401,Mac09256803,Hat14115103}, and so on. Therefore, it is highly desirable to develop a theoretical framework that can overcome these limitations and provide a more comprehensive understanding of open quantum systems.

The first limitation has been addressed by environmental quasiparticle theories \cite{Xu26021002}, such as dissipaton theory \cite{Yan14054105,Wan22170901,Wan22044102,Li23214110}, Lindblad pseudomode theory \cite{Gar972290,Ple20043058,Dal01053813,Luo23030316,Cir24033083,Hua26090403}, bexcitonics \cite{Che24204116}, and related approaches \cite{Xu26184116}. These theories retain environmental information through auxiliary degrees of freedom whose dynamics are coupled to those of the system. Dissipaton theory interprets the auxiliary density operators defined in the HEOM as irreducible moments of bath quasiparticles, providing access to bath observables and system--bath correlations. Pseudomode theory instead represents the structured environment through discrete auxiliary modes subject to damping, embedding the non-Markovian system dynamics in an enlarged dynamical description. The pseudomodes encode bath memory, while their joint evolution with the system retains correlations that are absent from the reduced system density operator alone. In particular, coupled Lindblad pseudomode constructions allow correlations between auxiliary modes to enter this representation \cite{Hua26090403}. These developments motivate an explicit operator description of the environmental degrees of freedom and their role in hierarchical dynamics.

The second limitation has also been explored within dissipaton theory. Introduced as a quasiparticle description of Gaussian environments \cite{Yan14054105}, dissipaton theory associates the exponential components of bath correlation functions with statistical quasiparticles, termed dissipatons. Yan et al. introduces the irreducible notation of dissipaton operators to define the dissipaton density operators, which retain information about the system and its correlations with the environment. The dynamics of these operators are governed by the dissipaton equations of motion (DEOM), constructed using generalized Wick contractions and a generalized diffusion equation \cite{Wan22170901}. Subsequent developments have enriched this framework through phase-space, thermofield, and embedded master-equation formulations \cite{Wan20041102,Wan22044102,Li23214110}. Beyond linear coupling, the generalized Wick contractions provide a systematic means of treating products of bath forces, leading to extensions of DEOM for nonlinear system--bath interactions, including quadratic coupling \cite{Xu17395,Xu18114103,Su24084104}. These advances establish dissipaton theory as a framework for investigating both reduced system dynamics and environmental responses in the presence of nonlinear interactions.

However, the dissipaton construction has been largely heuristic, with the dissipaton operators defined in terms of their correlation functions rather than as explicit operators in the bath space \cite{Jan25035111}. The generalized Wick contractions and diffusion equation are postulated to reproduce the HEOM structure, but their derivation from first principles remains incomplete. A more rigorous operator-based construction should be imposed to clarify the mathematical foundation of dissipaton theory and its extensions to nonlinear couplings.

In this work, motivated by dissipaton theory and the coupled pseudomode method, we give an exact and clear realization of the statistical quasiparticle representation of continuous baths, where non-hermitian creation/annihilation operators and complex eigenenergies are allowed. We also develop an ordered-moment approach that constructs hierarchical dynamics directly from dissipaton operators, without first introducing an influence functional. The main results are:
\begin{enumerate}[fullwidth]
    \item[(i)] We give dissipatons explicit definitions in terms of the underlying bath operators and derive their contraction rules and diffusion equation. This puts dissipaton theory on a firmer mathematical foundation.
    
    \item[(ii)] We derive hierarchical equations of motion from bath operator algebra and free evolution, bypassing the influence functional. Under the stated bath assumptions, the full hierarchies describe exact non-Markovian and nonperturbative dynamics.

    \item[(iii)] We provide a theoretical basis for treating nonlinear system--bath coupling. Products of bath forces are handled through operator ordering and Wick contractions. We demonstrate this approach for quadratic coupling and show how the same rules can be applied to higher-order couplings.
\end{enumerate}

This paper is organized as follows. Section~\ref{sec:discrete} introduces the ordered-moment approach for a discrete Gaussian bath, which serves as an instructive example of our theory. Section~\ref{sec:continuous} constructs explicit dissipaton operators for continuous baths under the stated spectral assumptions, derives their Langevin evolution and normal-ordered hierarchy, and develops the Liouville-space and single-index formulations. Section~\ref{sec:quadratic} treats quadratic system--bath coupling and extends the generalized Wick recursion to polynomial interactions of arbitrary finite order. Section~\ref{sec:conclusions} summarizes the results. Appendix~\ref{app:other-orderings} discusses the continuous-bath $G$ transformation and the relations between different orderings, with emphasis on the double-index hierarchy. Appendix~\ref{app:hermite} introduces generalized Hermite polynomials and shows how environmental moments and system--bath observables can be reconstructed from the ordered moments. Appendix~\ref{app:langevin} gives the coupled Lindblad representation of the bath evolution. Throughout this paper, we set $\hbar=1$ and define $\beta\equiv1/(k_BT)$, where $k_B$ is the Boltzmann constant and $T$ is the temperature.

\section{Discrete environment without dissipation}\label{sec:discrete}

In this section, we derive the hierarchical equations of motion for a discrete harmonic environment. The thermofield representation maps the thermal bath onto vacuum modes, whose normal-ordered moments define the dynamical variables. Their equations of motion follow directly from the total Liouville equation. This discrete setting provides an instructive starting point for the continuous-bath construction, where collective modes with decaying correlations replace the purely oscillatory modes considered here.

\subsection{System--bath model and thermofield representation}
Consider a system in contact with a harmonic bath, with the total Hamiltonian reading
\begin{equation}\label{linear}
 H_{\T}=H_{\tS}+H_{\B}+H_{\SB}.
\end{equation}
Here, $H_{\tS}$ describes the system, while the bath consists of $N$ independent harmonic modes,
\begin{equation}\label{discrete-bath-H}
 H_{\B}=\sum_{j=1}^N\omega_j\hat a_j^\dagger\hat a_j,
\end{equation}
where $\hat a_j^\dagger$ and $\hat a_j$ are the bosonic creation and annihilation operators of the $j$-th bath mode, respectively. The system--bath interaction takes the linear form
\begin{equation}\label{discrete-interaction-H}
 H_{\SB}=\hat Q\hat F,
\end{equation}
with $\hat Q=\hat Q^\dagger$ being a system operator and the bath interaction mode being
\begin{equation}\label{discrete-force}
 \hat F=\sum_{j=1}^N\frac{c_j}{\sqrt2}(\hat a_j+\hat a_j^\dagger).
\end{equation}
The frequencies $\omega_j$ are positive and the couplings $c_j$ are real. Here, for simplicity, we consider a single system--bath interaction channel, but the generalization to multiple channels is straightforward. 
The dynamics of the total system is governed by the Liouville--von Neumann equation, \Eq{Liouville}. The system--bath hybridizing process adopts the initial state being the direct product of an arbitrary system state and the bath canonical state, namely
\begin{align}
  \rho_{\T}(0) = \rho_{\tS}(0)\otimes \rho_{\B}^{\rm eq}(\beta) \equiv \rho_{\tS}(0)\otimes e^{-\beta H_{\B}}/Z_{\B}.
\end{align}
Here, $Z_{\B} \equiv {\rm tr}_{\B}\,e^{-\beta H_{\B}}$ is the bath canonical partition function.

To proceed, we introduce the thermofield decomposition to map the original canonical thermal state into an effective vacuum. Specifically, we expand \cite{Ume95}
\begin{align}\label{a2b}
  \hat a_j \equiv \sqrt{\bar{\mathscr{n}}_j+1}\hat b_j + \sqrt{\bar{\mathscr{n}}_j}\tilde b^\dagger_j,
\end{align}
such that the effective thermal vacuum state $|{\rm Vac}(\beta)\ra$ satisfies 
\begin{align}\label{vac}
  \hat b_j|{\rm Vac}(\beta)\ra = \tilde b_j|{\rm Vac}(\beta)\ra = 0
\end{align}
and the Heisenberg evolutions of $\hat b_j$ and $\tilde b_j$ are given by
\begin{align}\label{bt}
  \hat b_j^{\B}(t) = \hat b_j^{\B}(0)e^{-i\omega_jt},\quad \tilde b_j^{\B}(t) = \tilde b_j^{\B}(0)e^{i\omega_jt},
\end{align}
satisfying the commutation relations 
\begin{align}
  [\hat b_j,\hat b_{j'}^\dagger] = [\tilde b_j,\tilde b_{j'}^\dagger] = \delta_{jj'},\quad [\hat b_j,\tilde b_{j'}] = [\hat b_j,\tilde b_{j'}^\dagger] = 0.
\end{align}
In \Eq{a2b}, $\bar{\mathscr n}_j \equiv 1/(e^{\beta\omega_j}-1)$ is the average occupation number of the $j$-th bath mode. The vacuum is defined in the doubled space of the original bath Hilbert space, namely \cite{Ume95}
\begin{align}
  |\mathrm{Vac}(\beta)\ra \equiv \frac{1}{\sqrt{Z_{\B}}}\bigotimes_j\sum_{n_j=0}^\infty{e^{-\beta\omega_jn_j/2}}|n_j\ra\otimes|\tilde n_j\ra.
\end{align}
Here, the states $\{|n_j\ra\}$ are the eigenstates of the original bath $H_{\B}$, whereas $\{|\tilde n_j\ra\}$ are that of the auxiliary bath 
\begin{align}
  \tilde H_{\B} = -\sum_j\omega_j\tilde a_j^\dagger\tilde a_j
\end{align}
with
\begin{align}
  \tilde a_j = \sqrt{\bar{\mathscr{n}}_j+1}\tilde b_j + \sqrt{\bar{\mathscr{n}}_j}\hat b_j^\dagger.
\end{align}
Note that the auxiliary bath $\tilde H_{\B}$ actually is the time--reversed bath of $H_{\B}$, with basic eigenfrequencies being $\{-\omega_j\}$ \cite{Ume95}. Tracing over the auxiliary bath results in the original bath canonical state. 

As a result, the bath dynamics is governed by the Hamiltonian,
\begin{gather}
  \begin{split}
    {H}_{\B}^{\TF} = H_{\B} + \tilde H_{\B} = \sum_j\omega_j(\hat b^{\dagger}_j\hat b_j - \tilde b^{\dagger}_j\tilde b_j) \equiv \sum_{k=1}^{2N}\epsilon_k\hat d_k^\dagger\hat d_k,
  \end{split}
\end{gather}
where we introduce
\begin{align}
  \epsilon_k = \begin{cases}
    \omega_k, & k=1,\ldots,N,\\
    -\omega_{k-N}, & k=N+1,\ldots,2N,
  \end{cases}
\end{align}
and 
\begin{align}
  \hat d_k = \begin{cases}
    \hat b_k, & k=1,\ldots,N,\\
    \tilde b_{k-N}, & k=N+1,\ldots,2N.
  \end{cases}
\end{align}
We further recast the environment interaction mode as 
\begin{align}
  \hat F = \sum_jc_jx_j \equiv \sum_k\zeta_k(\hat d_k + \hat d_k^\dagger),
\end{align}
with
\begin{align}
  \zeta_k = \begin{cases}
    c_k\sqrt{(\bar{\mathscr n}_k + 1)/2}, & k=1,\ldots,N,\\
    c_{k-N}\sqrt{\bar{\mathscr n}_{k-N}/2}, & k=N+1,\ldots,2N.
  \end{cases}
\end{align}
 The thermal bath is thus represented by $2N$ vacuum modes with positive and negative frequencies, while its temperature dependence is contained in the coupling coefficients $\zeta_k$. The initial total state is mapped to 
\begin{align*}
  \rho_{\tS}(0)\otimes|{\rm Vac}(\beta)\ra\la{\rm Vac}(\beta)|.
\end{align*}
Hereafter, $\rho_{\T}$ denotes the density operator in the enlarged space and ${\rm tr}_{\B}$ includes the auxiliary bath trace. The total Hamiltonian in \Eq{linear} is correspondingly replaced by
\begin{align*}
  H_{\T} \to H_{\tS}+ H_{\B}^{\TF} +\hat Q\hat F,
\end{align*}
which recovers the original physical dynamics upon tracing out the auxiliary bath. 

\subsection{Ordered density operators and Wick contractions}

Define the ordered density operators (ODOs) as \cite{Su25_arxiv_2503_00297}
\begin{align}
  \rho_{\bm u,\bm v}(t) \equiv {\rm tr}_{\B}\bigg[ \prod_k\mathcal N \big(\hat d_k^{u_k}\hat d_k^{\dagger v_k} \big)\rho_{\T}(t) \bigg],
\end{align}
where $\mathcal N$ is the normal ordering defined as 
\begin{align}
  \mathcal N(\hat d_k^\dagger\hat d_k) = \mathcal N(\hat d_k\hat d_k^\dagger) = \hat d_k^\dagger\hat d_k
\end{align}
for all modes, and the index sets are denoted as $\bm u \equiv (u_1,\cdots,u_{2N})$ and $\bm v \equiv (v_1,\cdots,v_{2N})$ with $u_k, v_k = 0, 1, 2,\cdots$. For later use, we also define $\bm u_k^\pm \equiv(\cdots u_k \pm 1\cdots)$ and $\bm v_k^\pm \equiv(\cdots v_k \pm 1\cdots)$. The ODOs contain the correlated and entangled properties of the system and the bath, but represent the bath degrees of freedom into a set of ordered moments. Using \Eq{vac}, we obtain the initial condition: $\rho_{\bf 0,0}(0) = \rho_{\tS}(0)$ and others are zero. Combining \Eqs{Liouville} and (\ref{bt}), we obtain the equations of motion for $\rho_{\bm u,\bm v}$ as 
\begin{align}\label{EOM}
  \dot{\rho}_{\bm u,\bm v} &= -i[H_{\tS}, \rho_{\bm u,\bm v}] - i\sum_k(u_k - v_k)\epsilon_k\rho_{\bm u,\bm v} \nl
  &\quad\, -i \sum_k\zeta_k [\hat Q, \rho_{{\bm u}_k^+,\bm v} + \rho_{{\bm u},{\bm v}_k^+}] \nl
  &\quad\, -i \sum_k\zeta_k(u_k\hat Q\rho_{{\bm u}_k^-,\bm v} - v_k\rho_{{\bm u},{\bm v}_k^-}\hat Q).
\end{align}
In deriving \Eq{EOM}, we also use the Wick's contraction concerning with the normal ordering, namely
\begin{subequations}\label{discrete-wick}
\begin{align}
 \mathcal N(\hat d^u\hat d^{\dagger v})\hat d
 &=\mathcal N(\hat d^{u+1}\hat d^{\dagger v}) ,\\
 \mathcal N(\hat d^u\hat d^{\dagger v})\hat d^\dagger
 &=\mathcal N(\hat d^u\hat d^{\dagger v+1})+u\mathcal N(\hat d^{u-1}\hat d^{\dagger v}),\\
 \hat d\mathcal N(\hat d^u\hat d^{\dagger v})
 &=\mathcal N(\hat d^{u+1}\hat d^{\dagger v})+v\mathcal N(\hat d^u\hat d^{\dagger v-1}),\\
 \hat d^\dagger\mathcal N(\hat d^u\hat d^{\dagger v})
 &=\mathcal N(\hat d^u\hat d^{\dagger v+1}).
\end{align}
\end{subequations}
Equation (\ref{EOM}) is nothing but the double-index hierarchical equations of motion (HEOM) for the discretized bath \cite{Xu22230601} with correlation function given by
\begin{align}\label{ct}
  \la\hat F_{\B}(t)\hat F_{\B}(0)\ra_{\B} &= \sum_j \frac{c_j^2}{2}\Big[ e^{-i\omega_jt}\big( \bar{\mathscr{n}}_j + 1 \big) + e^{i\omega_jt}\bar{\mathscr n}_j \Big] \nl
  &\equiv \sum_k \zeta_k^2 e^{-i\epsilon_kt}.
\end{align}
However, our approach differs from the original construction of HEOM in the following aspects. Firstly, we define the dynamic variables $\rho_{\bm u,\bm v}$ without introducing the time ordering or the path integral representation of the influence functional. Secondly, the physical meanings of $\rho_{\bm u,\bm v}$ are straightforward, as the trace of one gives the corresponding ordered moments of bath modes. Thirdly and most importantly, deriving the equations of motion only utilizes the time evolution [\Eq{bt}] and Wick contraction of the normal ordering [\Eq{discrete-wick}], which largely overcomes the algebraic complexity of the influence functional, especially when nonlinear system--bath coupling is present; see \Sec{sec:quadratic}. It is worth noting that defining the ODOs in other ordering representations, such as anti-normal and Weyl orderings, also produce similar equations of motion as \Eq{EOM}. However, the normal ordering is the most convenient choice, since other orderings give non-zero initial conditions for $\rho_{\bm u,\bm v}$ with $\bm u = \bm v \neq \bf 0$.

Define the bath spectral density as
\begin{align}\label{Jw}
  J(\omega) = \frac{\pi}{2}\sum_jc_j^2\big[ \delta(\omega - \omega_j) - \delta(\omega + \omega_j) \big],
\end{align}
with the property $J(-\omega) = -J(\omega)$. Then we can recast \Eq{ct} as 
\begin{align}\label{fdt}
  \la\hat F_{\B}(t)\hat F_{\B}(0)\ra_{\B} = \frac{1}{\pi}\int_{-\infty}^\infty\!\!\ud\omega\,e^{-i\omega t}\frac{J(\omega)}{1 - e^{-\beta\omega}},
\end{align}
which is the well-known fluctuation--dissipation theorem. We see that the bath correlation function is completely determined by the bath spectral density $J(\omega)$, which is a continuous function in the thermodynamic limit. The discrete bath model can be viewed as a discretization of the continuous bath.

\section{Continuous environments: dissipaton representation}\label{sec:continuous}
We now turn to the continuous environment which means a bath whose physical oscillator frequencies range over a continuum, rather than the finite set in Sec.~\ref{sec:discrete}. For a single coupling channel, the microscopic total system Hamiltonian reads
\begin{gather}\label{continuum-physical-H}
\begin{split}
    H_{\T}&=H_{\tS}+H_{\B}+\hat Q\hat F,\\
 H_{\B}&=\int_0^\infty\!\!\ud\omega\,\omega\hat a^\dagger(\omega)\hat a(\omega),\\
 \hat F&=\int_0^\infty\!\!\ud\omega\,\frac{c(\omega)}{\sqrt2}
 [\hat a(\omega)+\hat a^\dagger(\omega)],
\end{split}
\end{gather}
where the bath creation and annihilation operators satisfy
\begin{align}
  [\hat a(\omega),\hat a^\dagger(\omega')]=\delta(\omega-\omega'), \quad [\hat a(\omega),\hat a(\omega')]=0,
\end{align}
and $c(\omega)$ is real, with the density of coupled modes absorbed into its definition. The field operators are distributions; the bath force is defined by integration against its coupling function. The continuum retains real oscillator frequencies and unitary free bath evolution; decay of correlations arises from their collective dephasing.

For a continuous bath, the spectral density [\Eq{Jw}] is assumed as a reasonably smooth function, being $J(\omega) = \pi c^2(\omega)/2$ and satisfies $J(\omega\to\infty) = 0$. It usually has a simple power--law behavior \cite{Wei21,Leg871}, 
\begin{align}
  J(\omega > 0) \propto \omega^af_{\rm c}(\omega;\omega_{\rm c})
\end{align}
and $J(-\omega) = -J(\omega)$, where $f_{\rm c}(\omega;\omega_{\rm c})$ is a cutoff function with $\omega_{\rm c}$ being the cutoff frequency.



One straightforward way to the dissipative dynamics is to evolute \Eq{EOM} with $N$ being a large number. The parameters $\{\zeta_k, \epsilon_k\}$ are obtained by discretizing the spectral density. However, this approach is computationally expensive, and thus nonrealistic and impractical. The other approach utilizes the fluctuation--dissipation theorem [\Eq{fdt}] to expand the bath correlation function in terms of a series of exponential functions, 
\begin{equation}\label{fexp}
  \begin{split}
    \la\hat F_{\B}(t)\hat F_{\B}(0)\ra_{\B} &\simeq \sum_{k=1}^K\eta_ke^{-\gamma_kt},\\
    \la\hat F_{\B}(0)\hat F_{\B}(t)\ra_{\B} &\simeq \sum_{k=1}^K\eta_{k}^*e^{-\gamma_k^*t},
  \end{split}
\end{equation}
with $t>0$. The amplitudes $\eta_k$ are generally complex. We take distinct decay rates $\gamma_k$ with $\operatorname{Re}\gamma_k>0$, real or occurring in complex-conjugate pairs.
A convergent sequence of exponential approximations can recover the original correlation as $K$ increases. The operator construction below assumes that the chosen finite expansion has a nonnegative spectrum and admits the stated stable factor; these properties must be checked separately for a fitted expansion. Within this representation, we use $C(t)$ for the represented correlation and assume $C(t=0)<\infty$. For a rational cutoff function, the exponential decomposition is evaluated via the Cauchy's residue theorem in contour integration. The integration via residues depends on not only the concrete form of the spectral density, but also the fractional decomposition of the bosonic function. For the latter, traditionally, people adopt the Mittag--Leffler decomposition, specifically named also as the Matsubara expansion.
Besides, the Pad\'{e} spectrum decomposition (PSD) \cite{Hu10101106,Hu11244106} can greatly decrease the number of decomposition terms of the bosonic function part for the same precision. By far, one of the most efficient and powerful expansions of \Eq{fexp} is the time-domain Prony fitting decomposition ($t$-PFD) \cite{Che22221102}, which fits the time correlation function with the minimum terms and is applied to arbitrary spectral density functions---including rational functions, exponential functions, step functions, etc. Related approaches include low-temperature noise decompositions \cite{Xu22230601} and minimal pole representations \cite{Zha25214111}; a comparison of exponential-decomposition methods is given in Ref.~\onlinecite{Tak24204105}. 

In the HEOM formalism, the exponential decomposition is a scheme for constructing closed hierarchical equations, since the derivative of an exponential function is still an exponential function. Here, in our theory, we treat each exponential term in \Eq{fexp} as a statistical quasiparticle, which we call a \emph{dissipaton} \cite{Yan14054105}. The dissipaton is a collective bath mode, with ${\rm Re}\,\gamma_k$ being its decay rate and ${\rm Im}\,\gamma_k$ being its oscillation frequency. When dissipaton was firstly introduced, Yan and coworkers \cite{Yan14054105,Wan22044102} presents its physical picture and the corresponding dissipaton algebra from the structure of the HEOM, without a clear definition of the dissipaton operator. In this section, we will give a rigorous definition of the dissipaton operator and proof the dissipaton algebra.

\subsection{Operator construction of dissipatons}\label{sec:dissipaton-construction}

To proceed, we also introduce the thermofield representation for the continuous bath, similar to the discrete case. The continuum thermofield representation combines the two branches of the doubled bath into a vacuum field on the real frequency axis. Denote the thermofield operator by $\hat d(\omega)$ with $\omega \in (-\infty,\infty)$, with the definition
\begin{align}
  \hat d(\omega) = \begin{cases}
    \hat b(\omega), & \omega > 0,\\
    \tilde b(-\omega), & \omega < 0,
  \end{cases}
\end{align}
Here, 
\begin{subequations}
  \begin{align}
    \hat b(\omega) &\equiv \sqrt{\bar{\mathscr{n}}(\omega)+1}\hat a(\omega) - \sqrt{\bar{\mathscr{n}}(\omega)}\tilde a^\dagger(\omega),\\
    \tilde b(\omega) &\equiv \sqrt{\bar{\mathscr{n}}(\omega)+1}\tilde a(\omega) - \sqrt{\bar{\mathscr{n}}(\omega)}\hat a^\dagger(\omega),
  \end{align}
\end{subequations}
with $\tilde{a}(\omega)$ being the auxiliary bath operator and $\bar{\mathscr n}(\omega) \equiv 1/(e^{\beta\omega}-1)$ being the average occupation number of the bath mode with frequency $\omega$. Thus, the thermofield operator satisfies
\begin{equation}\label{continuous-vacuum}
 [\hat d(\omega),\hat d^\dagger(\omega')]=\delta(\omega-\omega'),
 \quad \hat d(\omega)|{\rm Vac}(\beta)\ra=0.
\end{equation}
The field operators are understood as distributions. Consequently, the bath generator in the thermofield representation is 
\begin{align}
  H_{\B}^{\TF}=\int_{-\infty}^\infty\!\!\ud\omega\,\omega\,\hat d^\dagger(\omega)\hat d(\omega),
\end{align}
including the negative-frequency auxiliary branch. Hereafter, we will omit the superscript $\rm TF$ for brevity when no ambiguity arises.
For the channel coupled to $\hat Q$, the freely evolving bath force can be written as
\begin{align}\label{continuous-force-vacuum}
 \hat F_{\B}(t)&=\hat F^{-}_{\B}(t)+\hat F^{+}_{\B}(t),\\
 \hat F^{-}_{\B}(t)&=\frac1{\sqrt{\pi}}\int_{-\infty}^\infty\!\!\ud\omega\,e^{-i\omega t}
 \sqrt{C(\omega)}\hat d(\omega),
\end{align}
with $\hat F^{+}_{\B}(t)=[\hat F^{-}_{\B}(t)]^\dagger$ and 
\begin{align}\label{Sw}
  C(\omega)\equiv\frac{1}{2}\int_{-\infty}^\infty \ud t\, e^{i\omega t}C(t) = {\rm Re}\int_0^\infty\!\! \ud t\, e^{i\omega t}C(t)
\end{align}
being the Fourier transform of the bath correlation function. It is easy to verify that $C(\omega)$ is real. From the fluctuation--dissipation theorem [\Eq{fdt}], we have 
\begin{align}\label{continuous-spectrum}
  C(\omega) = \frac{J(\omega)}{1 - e^{-\beta\omega}}  \geq 0.
\end{align}
The vacuum contraction gives
\begin{align}
  &\quad\,\la\hat F_{\B}(t)\hat F_{\B}(0)\ra_{\B} \nl
  &= \frac{1}{\pi}\int_{-\infty}^\infty\!\!\ud\omega\int_{-\infty}^\infty\!\!\ud\omega' e^{-i\omega t}\sqrt{C(\omega)C(\omega')}\la\hat d(\omega)\hat d^\dagger(\omega')\ra_{\B}\nl
  &= \frac{1}{\pi}\int_{-\infty}^\infty\!\!\ud\omega\,e^{-i\omega t}C(\omega) = C(t),
\end{align}
where the last identity is the inverse Fourier transform of \Eq{Sw}. 

To resolve \Eq{continuous-force-vacuum} into exponential modes at the operator level, we need a corresponding decomposition of the coupling amplitude. Although under the exponential decomposition the spectrum 
\begin{align}
  C(\omega) = {\rm Re}\,\sum_k \frac{\eta_k}{{\gamma_k-i\omega}}
\end{align}
is rational, its nonnegative square root $\sqrt{C(\omega)}$ is generally not rational and does not directly admit a finite partial-fraction expansion. The spectrum, however, fixes only the modulus of this amplitude: its phase can be absorbed into the continuum annihilation field without changing the bath force or its correlations. We use this freedom to introduce a stable rational spectral factor,
\cite{Mueller26}
\begin{align}\label{stable-spectral-factor}
 C(\omega)&=|h(\omega)|^2,\quad
 h(\omega)=\sum_{k=1}^{K}\frac{r_k}{\gamma_k-i\omega},
\end{align}
with $\{r_k\}$ being complex coefficients.
Here, stability means that all poles of $h(\omega)$ lie in the
lower half of the complex frequency plane, since
$\operatorname{Re}\gamma_k>0$. Once a stable spectral factor $h(\omega)$ has been chosen, the coefficients $r_k$ are obtained by its partial-fraction expansion \cite{Mueller26}. They are generally complex and depend on the phase choice of the spectral factor, whereas the bath spectrum $C(\omega)=|h(\omega)|^2$ is unchanged. Each pole in \Eq{stable-spectral-factor} therefore corresponds to a causal, decaying exponential,
\begin{equation}
 \frac{1}{\gamma_k-i\omega}
 =\int_0^\infty\!\ud t\,e^{-\gamma_k t}e^{i\omega t}.
\end{equation}
Consequently, $h(\omega)$ is the frequency-domain amplitude of a causal filter with memory kernel $\sum_kr_k e^{-\gamma_k t}$. Physically, this filter converts a delta-correlated vacuum input into the colored bath force with spectrum $C(\omega)$, as made explicit below. The spectral factor thus connects the bath spectrum to a realization in terms of exponential memory modes. It describes a representation of the collective bath fluctuations; its poles are not the frequencies of individual physical bath oscillators. The coefficients $r_k$ are determined once a spectral factor of $C(\omega)$ is chosen. This factor need not be unique, because $C(\omega)$ specifies its modulus but not its phase.

To implement this choice without changing the force operator,
introduce the local--phase transformed field
\begin{align}\label{continuum-phase}
  \tilde{d}(\omega)\equiv e^{i\varphi(\omega)}\hat d(\omega),
  \quad \text{with}\quad e^{i\varphi(\omega)}\equiv \frac{\sqrt{C(\omega)}}{h(\omega)}.
\end{align}
Since $C(\omega)=|h(\omega)|^2$, the prefactor has unit modulus
wherever $C(\omega)>0$. At zeros of $C(\omega)$, where the coupling vanishes, we define it to be unity.  Hence, this transformation preserves the canonical commutators and the thermofield vacuum. We define the dissipaton annihilation operator by
\begin{equation}\label{dissipaton-operator-definition}
 {\hat f_k^{\B}(t) \equiv e^{i H_{\B}t}\hat f_ke^{-i H_{\B}t} \equiv \frac1{\sqrt{\pi}}\int_{-\infty}^\infty\!\!\ud\omega\,e^{-i\omega t}
 \frac{r_k}{\gamma_k-i\omega}\tilde d(\omega)
 .}
\end{equation}
Hereafter, $\hat f_k = \hat f_k^{\B}(0)$, and $\hat f_k^\dagger$ denotes its Hermitian conjugate. 
Thus, \Eq{dissipaton-operator-definition} defines a smeared bosonic annihilation operator on the common dense domain of finite-particle vectors in the thermofield Fock space. 
Summing over $k$ and using \Eqs{stable-spectral-factor} and (\ref{continuum-phase}), we obtain
\begin{equation}\label{continuous-dissipaton-force}
 \hat F^{-}_{\B}(t)=\sum_k\hat f_k^{\B}(t),\quad
 \hat F_{\B}(t)=\sum_k\bigl[\hat f_k^{\B}(t)+\hat f_k^{\dagger\B}(t)\bigr].
\end{equation}
Note that the bath coupling amplitudes are included in $\hat f_k$. 

Since \Eq{dissipaton-operator-definition} contains only vacuum annihilation fields, the thermofield property is retained, with
\begin{equation}\label{dissipaton-vacuum-property}
 \hat f_k^{\B}(t)|{\rm Vac}(\beta)\ra=0,
\end{equation}
resulting in 
\begin{align}
  \la\hat f_i^\dagger\hat f_j\ra_{\B}
 =\la\hat f_i\hat f_j\ra_{\B}=0.
\end{align}
The transformation to the exponential modes is generally nonorthogonal. Their equal-time commutators are given by 
\begin{align}\label{dissipaton-gram}
 [\hat f_i,\hat f_j^\dagger]&=M_{ij},\quad
 [\hat f_i,\hat f_j]=0,
\end{align}
with
\begin{align}
M_{ij}&=\frac1{\pi}\int_{-\infty}^\infty\ud\omega\,
 \frac{r_i r_j^*}
 {(\gamma_i-i\omega)(\gamma_j^*+i\omega)}=\frac{2r_i r_j^*}{\gamma_i+\gamma_j^*}.
\end{align}
In particular, $\la\hat f_i\hat f_j^\dagger\ra_{\B}=M_{ij}$ and $M_{ji}=M_{ij}^*$. The matrix $\mathbf M$ is positive semidefinite, since for any complex vector $\bm a=(a_1,\dots,a_K)^T$,
\begin{equation}
 \bm a^\dagger \mathbf M\bm a
 =2\int_0^\infty\!\!\ud t\,
 \left|\sum_k a_k^*r_k e^{-\gamma_k t}\right|^2
 \geq0.
\end{equation}
For distinct $\gamma_k$ and nonzero $r_k$, the exponential
kernels are linearly independent, so $\mathbf M$ is positive definite.

The dissipatons share a vacuum but are not independent canonical modes. Their nonorthogonal algebra, rather than a canonical change of basis, will be used to construct the hierarchy. For $t\geq0$, we have
\begin{equation}\label{dissipaton-cross-correlation}
 \la\hat f_i^{\B}(t)\hat f_j^\dagger\ra_{\B}
 =e^{-\gamma_i t}M_{ij}.
\end{equation}
It follows from \Eq{continuous-dissipaton-force} that
\begin{equation}\label{dissipaton-residue-rowsum}
 C(t)=\sum_i e^{-\gamma_i t}\sum_jM_{ij},
\end{equation}
leading to 
\begin{align}
  \eta_i=\sum_jM_{ij}.
\end{align}
The complex amplitude $\eta_i$ is therefore a sum of cross correlations, while the individual variance $M_{ii}=|r_i|^2/\operatorname{Re}\gamma_i$ remains real and positive. 

\subsection{Generalized Langevin equation for dissipatons}

The dissipative character of the exponential modes can be seen directly from their free evolution. Define 
\begin{equation}\label{dissipaton-input}
 \hat\xi(t)\equiv\frac1{\sqrt{\pi}}\int_{-\infty}^\infty\!\!\ud\omega\,e^{-i\omega t}
 \tilde d(\omega),
\end{equation}
satisfying
\begin{align}
  [\hat\xi(t),\hat\xi^\dagger(t')]=2\delta(t-t').
\end{align}
Then \Eq{dissipaton-operator-definition} can be recast as 
\begin{equation}\label{dissipaton-causal-filter}
 \hat f_k^{\B}(t)=r_k\int_{-\infty}^{t}\!\!\ud\tau\,
 e^{-\gamma_k(t-\tau)}\hat\xi(\tau).
\end{equation}
Differentiating with respect to $t$ gives the generalized Langevin equation for the dissipaton operators,
\begin{equation}\label{dissipaton-langevin}
 \frac{\ud}{\ud t}\hat f_k^{\B}(t) 
 =-\gamma_k\hat f_k^{\B}(t)+r_k\hat\xi(t).
\end{equation}
In fact, $\hat\xi(t)$ can be seen as a Gaussian vacuum white-noise input with zero mean and correlations
\begin{subequations}
  \begin{align}
 \langle\hat\xi(t)\hat\xi^\dagger(t')\rangle_{\B}
 &=2\delta(t-t'),\\
 \langle\hat\xi^\dagger(t)\hat\xi(t')\rangle_{\B}
 &=\langle\hat\xi(t)\hat\xi(t')\rangle_{\B}=0.
\end{align}
\end{subequations}
Equation~\eqref{dissipaton-langevin} admits a coupled Lindblad representation for the collective bath modes after tracing out the common vacuum input \cite{Gar853761}. Appendix~\ref{app:langevin} derives its Hamiltonian and jump operator and verifies the corresponding free moment dynamics. 


\subsection{Wick contractions and hierarchical equations}\label{sec:continuous-hierarchy}
Similar as the discrete case, defining the normal ordering of the dissipaton operators is the key to construct hierarchical equations. The nondiagonal commutators require a global ordering of the multimode moment,
\begin{equation}\label{continuous-normal-product}
 \mathcal N(\bm u,\bm v)
 \equiv\mathcal N\!\left(\prod_k\hat f_k^{u_k}\hat f_k^{\dagger v_k}\right)
 =\left(\prod_k\hat f_k^{\dagger v_k}\right)
  \left(\prod_k\hat f_k^{u_k}\right).
\end{equation}
All creation operators are placed to the left of all annihilation operators, including those of different modes. The corresponding ODOs in the dissipaton representation are
\begin{equation}\label{continuous-odo}
 \rho_{\bm u,\bm v}(t)
 \equiv {\rm tr}_{\B}\left[\mathcal N(\bm u,\bm v)\rho_{\T}(t)\right],
\end{equation}
where $\bm u,\bm v$ now have $K$ components. These ODOs are system operators weighted by bath moments, with $\rho_{\mathbf0,\mathbf0}=\rho_{\tS}$.

Using \Eq{dissipaton-gram} in the globally ordered product gives
\begin{subequations}\label{continuous-wick}
\begin{align}
 \mathcal N(\bm u,\bm v)\hat f_j
 &=\mathcal N(\bm u_j^+,\bm v),\\
 \mathcal N(\bm u,\bm v)\hat f_j^\dagger
 &=\mathcal N(\bm u,\bm v_j^+) + \sum_i u_iM_{ij}\mathcal N(\bm u_i^-,\bm v),\\
 \hat f_j\mathcal N(\bm u,\bm v)
 &=\mathcal N(\bm u_j^+,\bm v) +\sum_i v_iM_{ji}\mathcal N(\bm u,\bm v_i^-),\\
 \hat f_j^\dagger\mathcal N(\bm u,\bm v)
 &=\mathcal N(\bm u,\bm v_j^+).
\end{align}
\end{subequations}
The second identity follows from $[\hat f_i^{u_i},\hat f_j^\dagger]=u_iM_{ij}\hat f_i^{u_i-1}$, while the third follows by commuting $\hat f_j$ through the creation block. The other two insertions are already normally ordered. 
The system couples to the total force $\hat F=\sum_k(\hat f_k+\hat f_k^\dagger)$. Summing \Eqs{continuous-wick} and using $\sum_jM_{ij}=\eta_i$ and $\sum_jM_{ji}=\eta_i^*$, we obtain
\begin{subequations}\label{continuous-force-insertions}
\begin{align}
 \mathcal N(\bm u,\bm v)\hat F
 &=\sum_k\bigl[\mathcal N(\bm u_k^+,\bm v)
             +\mathcal N(\bm u,\bm v_k^+)\bigr]\nl
 &\quad+\sum_k u_k\eta_k\mathcal N(\bm u_k^-,\bm v),\\
 \hat F\mathcal N(\bm u,\bm v)
 &=\sum_k\bigl[\mathcal N(\bm u_k^+,\bm v)
             +\mathcal N(\bm u,\bm v_k^+)\bigr]\nl
 &\quad+\sum_k v_k\eta_k^*\mathcal N(\bm u,\bm v_k^-).
\end{align}
\end{subequations}
The distinction between the two contraction coefficients therefore follows from the ordinary left and right multiplication of the bath force. Although individual dissipaton insertions involve the full matrix $\bf M$, the interaction vertices depend only on its row and column sums.

The free bath contribution follows from an operator identity over a finite forward time interval. In the bath interaction picture, denote 
\begin{align}
  \rho_{\T}^{I}(t)\equiv e^{i H_{\B}t}\rho_{\T}(t)e^{-i H_{\B}t}.
\end{align}
Splitting the integral in \Eq{dissipaton-causal-filter} at $t$ gives the exact identity for $\Delta t>0$
\begin{align}\label{continuous-finite-step}
 \hat f_k^{\B}(t+\Delta t)&=e^{-\gamma_k\Delta t}\hat f_k^{\B}(t)+\hat\nu_{k}(t;\Delta t),
\end{align}
with
\begin{align}
  \hat\nu_{k}(t;\Delta t)&\equiv r_k\int_t^{t+\Delta t}\!\!\ud \tau\,
 e^{-\gamma_k(t+\Delta t-\tau)}\hat\xi(\tau).
\end{align}
For $0\le\tau\le t$, the field commutator in \Eq{dissipaton-input} yields
\begin{align}\label{continuous-future-commutator}
\! [\hat\nu_{k}(t;\Delta t),\hat f_l^{\dagger\B}(\tau)]
 ={}&2r_kr_l^*\int_t^{t+\Delta t}\!\!\!\!\!\ud t_2\int_{-\infty}^{\tau}\!\!\ud t_1\,
 e^{-\gamma_k(t+\Delta t-t_2)}\nl
 &\times e^{-\gamma_l^*(\tau-t_1)}\delta(t_2-t_1)=0.
\end{align}
Consequently, $\hat\nu_{k}(t;\Delta t)$ commutes with $H_{\tS}+\hat Q\hat F_{\B}(\tau)$ throughout the propagation interval $[0,t]$, and hence with the propagator $U_I(t,0) = \mathcal T_+e^{-i\int_0^t\ud \tau [H_{\tS}+\hat Q\hat F_{\B}(\tau)]}$. 
Since $\hat\nu_{k}(t;\Delta t)|{\rm Vac(\beta)}\rangle=0$ and $\rho_{\T}^I(0) = \rho_{\tS}(0)\otimes |\rm Vac(\beta)\ra\la \rm Vac(\beta)|$, the specified factorized initial state gives
\begin{align}\label{continuous-future-vacuum}
 \hat\nu_{k}(t;\Delta t)\rho_{\T}^{I}(t)
 &=U_I(t,0)\hat\nu_{k}(t;\Delta t)\rho_{\T}^{I}(0)U_I^\dagger(t,0)=0.
\end{align}

Let $\mathcal N_{\B}(t;\bm u,\bm v)$ denote the normal product of the freely evolving dissipatons, namely
\begin{align}\label{continuous-normal-product-time}
  \mathcal N_{\B}(t;\bm u,\bm v) \equiv \mathcal N\!\left[\prod_k\hat f_k^{\B}(t)^{u_k}\hat f_k^{\dagger\B}(t)^{v_k}\right],
\end{align}
and set $\gamma_{\bm u,\bm v} \equiv\sum_k(u_k\gamma_k+v_k\gamma_k^*)$. Then we have 
\begin{align}
  \rho_{\bm u,\bm v}(t) = {\rm tr}_{\B}\left[\mathcal N_{\B}(t;\bm u,\bm v)\rho_{\T}^{I}(t)\right].
\end{align}
Substituting \Eq{continuous-finite-step} into \Eq{continuous-normal-product-time} leaves all additional creation operators to the left of the additional annihilation operators. Each term containing an additional annihilation operator therefore vanishes against $\rho_{\T}^{I}(t)$ by \Eq{continuous-future-vacuum}; terms containing only additional creation operators vanish by cyclicity of the bath trace. Hence,
\begin{equation}\label{continuous-normal-finite-trace}
 {\rm tr}_{\B}[\mathcal N_{\B}(t+\Delta t;\bm u,\bm v)\rho_{\T}^{I}(t)]
 =e^{-\Delta t\gamma_{\bm u,\bm v}}\rho_{\bm u,\bm v}(t).
\end{equation}
The density operator is held fixed at $t$ in this identity. The free bath contribution is defined by its \emph{right derivative},
\begin{align}\label{continuous-normal-diffusion}
 &\quad\,
 \left.\partial_t^+\rho_{\bm u,\bm v}\right|_{\B}\nl
 &\equiv\lim_{\Delta t\to0^+}\frac{1}{\Delta t}{\rm tr}_{\B}\bigl[
 \bigl(\mathcal N_{\B}(t+\Delta t;\bm u,\bm v) - \mathcal N_{\B}(t;\bm u,\bm v)\bigr)\rho_{\T}^{I}(t)\bigr]\nl
 &=-\gamma_{\bm u,\bm v}\rho_{\bm u,\bm v}(t).
\end{align}
This is the generalized diffusion equation for normally ordered moments. The restriction $\Delta t\to0^+$ is essential to the present proof: the time evolution in $(t-\Delta t,t]$ has already been accessible to the interaction and need not annihilate $\rho_{\T}^I(t)$.

The remaining contribution follows from
$\dot\rho_{\T}^{I}=-i[H_{\tS}+\hat Q\hat F_{\B}(t),\rho_{\T}^{I}]$.
Dots below denote the forward time derivatives. Combining the Liouville--von Neumann equation with \Eqs{continuous-normal-diffusion} and (\ref{continuous-force-insertions}), we obtain
\begin{align}\label{continuous-normal-heom}
  \dot\rho_{\bm u,\bm v} &= -i[H_{\tS},\rho_{\bm u,\bm v}] - \sum_k(u_k\gamma_k+v_k\gamma_k^*)\rho_{\bm u,\bm v}\nl
  &\quad\, -i\sum_k[\hat Q,\rho_{\bm u_k^+,\bm v}+\rho_{\bm u,\bm v_k^+}]\nl
  &\quad\, -i\sum_k \Big( u_k\eta_k\hat Q\rho_{\bm u_k^-,\bm v} - v_k\eta_k^*\rho_{\bm u,\bm v_k^-}\hat Q \Big).
\end{align}
Equation~(\ref{continuous-normal-heom}) is exactly the double-index HEOM \cite{Xu22230601}. For the factorized thermal preparation, the thermofield vacuum gives
\begin{equation}\label{continuous-initial}
 \rho_{\bm u,\bm v}(0)
 =\delta_{\bm u,\mathbf0}\delta_{\bm v,\mathbf0}\rho_{\tS}(0).
\end{equation}

Actually, we can also derive different hierarchical equations by using other orderings, such as the Weyl and anti-normal orderings. However, the normal ordering is the most convenient one: (i) the initial condition is simple and (ii) the final equations of motion are independent of the choice of the factors $\{r_k\}$. See Appendix~\ref{app:other-orderings} for details.

\subsection{Liouville space representation of dissipatons}

The dissipaton operators constructed above have the nonorthogonal commutator matrix $\mathbf M$. Consequently, inserting an individual $\hat f_k^\dagger$ into a normally ordered moment produces contractions with several modes, as shown in \Eq{continuous-wick}. To resolve these contractions into separate mode contributions, we introduce operators dual to $\{\hat f_k\}$. For distinct poles and nonzero residues, $\mathbf M$ is positive definite and hence invertible, so we define 
\begin{align}\label{dual-dissipaton-definition}
  \hat g_k \equiv \sum_j (\mathbf M^{-1})_{kj}\hat f_j.
\end{align}
Since $\mathbf M^{-1}$ is Hermitian, $\hat g_k^\dagger=\sum_j(\mathbf M^{-1})_{jk}\hat f_j^\dagger$. The duality is expressed by
\begin{subequations}\label{eq:dissipaton-commutator}
  \begin{align}
  [\hat g_k, \hat f_{k'}^\dagger] = \sum_j \mathbf M^{-1}_{kj}\mathbf M_{jk'} = \delta_{kk'}
\end{align}
and
\begin{align}
  [\hat f_k, \hat g_{k'}^\dagger] = \sum_j \mathbf M_{kj}\mathbf M^{-1}_{jk'} = \delta_{kk'}.
\end{align}
\end{subequations}
These relations isolate a single occupation index: commuting $\hat g_k^\dagger$ through the annihilation block lowers $u_k$, while commuting $\hat g_k$ through the creation block lowers $v_k$. The dual operators therefore provide explicit maps for the contractions in Liouville space. Using $\eta_k = \sum_l M_{kl}$, we also obtain
\begin{align}
  \sum_k\eta_k\hat g_k^\dagger &= \sum_{kl}\eta_k(\mathbf M^{-1})_{lk}\hat f_l^\dagger \nl 
  &= \sum_{klj} (\mathbf M)_{kj}(\mathbf M^{-1})_{lk}\hat f_l^\dagger \nl
  &= \sum_k \hat f_k^\dagger.
\end{align}
Its Hermitian conjugate gives $\sum_k\eta_k^*\hat g_k=\sum_k\hat f_k$. These identities ensure that resolving the contractions with dual operators preserves the total bath force. For convenience, we define the superoperators with respect to an arbitrary system operator $\hat A$ as
\begin{subequations}\label{operator-superoperators}
  \begin{align}
  \hat A^>\hat O &\equiv \hat A\hat O,\\
  \hat A^<\hat O &\equiv \hat O\hat A,\\
  \hat A^\times\hat O &\equiv [\hat A,\hat O]
\end{align}
\end{subequations}
Using the left, right, and commutator actions defined in \Eq{operator-superoperators}, we introduce the dissipaton superoperators
\begin{subequations}
  \begin{align}
    \mathscr f_k^{->} &\equiv \hat f_k^>,\\
    \mathscr f_k^{+>} &\equiv \hat f_k^{\dagger <} + \eta_k\hat g_k^{\dagger\times}, \\
    \mathscr f_k^{-<} &\equiv \hat f_k^> - \eta_k^*\hat g_k^\times, \\
    \mathscr f_k^{+<} &\equiv \hat f_k^{\dagger <}.
  \end{align}
\end{subequations}
The pure multiplication terms generate normally ordered moments of the next tier, while the commutator terms select the lower-tier contractions. These superoperators are explicit linear maps on bath operator space; their $>$ and $<$ labels distinguish the two insertion branches and do not imply ordinary left and right multiplication by the same individual Hilbert-space operator. Their sums nevertheless recover the physical force actions exactly:\begin{subequations}
  \begin{align}
  \sum_k(\mathscr f_k^{->} + \mathscr f_k^{+>}) &= \sum_k(\hat f_k^> + \hat f_k^{\dagger >}) = \hat F^>,\\
  \sum_k(\mathscr f_k^{-<} + \mathscr f_k^{+<}) &= \sum_k(\hat f_k^< + \hat f_k^{\dagger <}) = \hat F^<.
\end{align}
\end{subequations}
The duality relations also yield the following nonzero commutators, 
\begin{subequations}
  \begin{align}
    [\mathscr f_k^{->},\mathscr f_{k'}^{+>}] &= \delta_{kk'}\eta_{k}, \quad [\mathscr f_k^{-<},\mathscr f_{k'}^{+<}] = -\delta_{kk'}\eta_k^*, \\
    [\mathscr f_k^{+>},\mathscr f_{k'}^{-<}] &= -\delta_{kk'}(\eta_k-\eta_k^*).
  \end{align}
\end{subequations}
All other commutators vanish. In particular, superoperators with different mode indices commute, although the two branches of the same mode need not commute when $\eta_k$ is complex.

To obtain their action on the ODOs, use cyclicity of the bath trace to transfer each commutator to $\mathcal N(\bm u,\bm v)$. The dual relations give $[\mathcal N(\bm u,\bm v),\hat g_k^\dagger]=u_k\mathcal N(\bm u_k^-,\bm v)$ and $[\hat g_k,\mathcal N(\bm u,\bm v)]=v_k\mathcal N(\bm u,\bm v_k^-)$. The resulting contractions are
\begin{subequations}\label{eq:dissipaton_algebra_0}
\begin{align}
  {\rm tr}_{\B}[\mathcal N(\bm u,\bm v)\mathscr f_k^{->}\rho_{\T}] &=  \rho_{\bm u_k^+,\bm v},\\
  {\rm tr}_{\B}[\mathcal N(\bm u,\bm v)\mathscr f_k^{+>}\rho_{\T}] &=  \rho_{\bm u,\bm v_k^+} + u_k\eta_k \rho_{\bm u_k^-,\bm v},\\
  {\rm tr}_{\B}[\mathcal N(\bm u,\bm v)\mathscr f_k^{-<}\rho_{\T}] &=  \rho_{\bm u_k^+,\bm v} + v_k\eta_k^* \rho_{\bm u,\bm v_k^-},\\
  {\rm tr}_{\B}[\mathcal N(\bm u,\bm v)\mathscr f_k^{+<}\rho_{\T}] &=  \rho_{\bm u,\bm v_k^+}.
\end{align}
\end{subequations}
The absence of a contraction in the first and fourth lines follows from placing annihilation operators at the right and creation operators at the left of the normal product. The other two lines contain only the selected mode, with coefficients $\eta_k$ and $\eta_k^*$. Thus, the dual construction converts the cross-mode contractions of the Hilbert-space representation into mode-resolved insertions without altering their summed physical action.

\subsection{Single-index hierarchy}

The two sets of occupation indices can be combined when $\{\gamma_k\}$ are real or conjugated pairs. Let $\bar k$ denote the partner of $k$, with $\gamma_{\bar k}=\gamma_k^*$ and $\bar{\bar k}=k$; a real one is its own partner. The combinations
\begin{equation}\label{single-dissipatons}
 \hat\phi_k\equiv\hat f_k+\hat f_{\bar k}^\dagger
\end{equation}
combine two components with the same free drift rate. Their Langevin evolution still contains the vacuum input; the common drift rate closes the evolution of the projected normal moments through \Eq{continuous-normal-diffusion}. Moreover, $\hat\phi_k^\dagger = \hat\phi_{\bar k}$ and $\hat F = \sum_k\hat\phi_k$.

Define the single-index ODOs as 
\begin{align}\label{single-index-odo}
  \rho_{\bm n}(t) \equiv {\rm tr}_{\B}\bigg[ \mathcal N\bigg(\prod_k\hat\phi_k^{n_k}\bigg)\rho_{\T}(t) \bigg].
\end{align}
Here $\bm n \equiv (n_1,\cdots,n_K)$ and the ordering is still defined with respect to the original dissipaton operators, $\{\hat f_k\}$ and $\{\hat f_k^\dagger\}$. Explicitly, we have 
\begin{align}
  \rho_{\bm n} =\sum_{\bm m\leq\bm n}
 \binom{\bm n}{\bm m}
 \rho_{\bm m,\overline{\bm n-\bm m}},
\end{align}
where $(\overline{\bm m})_k\equiv m_{\bar k}$. Each term chooses $m_k$ annihilation factors and $n_k-m_k$ creation factors from $\hat\phi_k^{n_k}$ before restoring global normal ordering. Its free drift is $\sum_k n_k\gamma_k$, independent of this choice. Projecting \Eq{continuous-normal-heom} therefore gives the single-index HEOM,
\begin{align}\label{single-normal-heom}
  \dot\rho_{\bm n}
  ={}&-i[H_{\tS},\rho_{\bm n}]
  -\sum_k n_k\gamma_k\rho_{\bm n} -i\sum_k[\hat Q,\rho_{\bm n_k^+}]\nl
  &-i\sum_k n_k(\eta_k\hat Q\rho_{\bm n_k^-}-\eta_{\bar k}^*\rho_{\bm n_k^-}\hat Q).
\end{align}
The two lower-tier coefficients retain the distinction between the forward and reverse bath correlations in \Eq{fexp}. To express this hierarchy directly through dissipaton superoperators, write the single-index normal product as 
\begin{align}
  \mathcal N(\bm n) \equiv \mathcal N\bigg(\prod_k\hat\phi_k^{n_k}\bigg).
\end{align}
Both insertion branches must raise the same index $n_k$. This is achieved by the common term $\hat f_k^>+\hat f_{\bar k}^{\dagger<}$, whose two contributions add one normally ordered $\hat\phi_k$. To lower this index, the left branch contracts $\hat f_k$ with $\hat g_k^\dagger$, whereas the right branch contracts $\hat f_{\bar k}^\dagger$ with $\hat g_{\bar k}$. We accordingly define
\begin{subequations}\label{single-insertion-maps}
  \begin{align}
  \mathscr f_k^> &\equiv \hat f_k^> + \hat f_{\bar k}^{\dagger <} + \eta_k\hat g_k^{\dagger\times}, \\
  \mathscr f_k^< &\equiv \hat f_k^> + \hat f_{\bar k}^{\dagger <} - \eta_{\bar k}^*\hat g_{\bar k}^\times,
\end{align}
\end{subequations}
These maps are linear combinations of the double-index insertion superoperators above. They generally differ from ordinary multiplication by $\hat\phi_k$, but satisfy
\begin{align}
  \!\!\!\![\mathscr f_k^>,\mathscr f_{k'}^>] = [\mathscr f_k^<,\mathscr f_{k'}^<] = 0, \quad [\mathscr f_k^>,\mathscr f_{k'}^<] = -\delta_{kk'}(\eta_k-\eta_{\bar k}^*)
\end{align}
and 
\begin{align}\label{single-force-decomposition}
  \hat F^{\gtrless} = \sum_k\mathscr{f}_k^{\gtrless}.
\end{align}
In particular, the nonzero same-mode cross commutators sum to zero because $\sum_k\eta_k$ is real, consistently with $[\hat F^>,\hat F^<]=0$. For the normal products, \Eq{eq:dissipaton-commutator} gives both $[\mathcal N(\bm n),\hat g_k^\dagger]=n_k\mathcal N(\bm n_k^-)$ and $[\hat g_{\bar k},\mathcal N(\bm n)]=n_k\mathcal N(\bm n_k^-)$. Together with the common raising term, these identities yield 
\begin{subequations}\label{eq:dissipaton_algebra}
  \begin{align}
    {\rm tr}_{\B}[\mathcal N(\bm n)\mathscr{f}_k^>\rho_{\T}] &= \rho_{\bm n_k^+} + n_k\eta_k\rho_{\bm n_k^-}, \\
    {\rm tr}_{\B}[\mathcal N(\bm n)\mathscr{f}_k^<\rho_{\T}] &= \rho_{\bm n_k^+} + n_{k}\eta_{\bar k}^*\rho_{\bm n_{k}^-}.
  \end{align}
\end{subequations}
Equation~\eqref{single-force-decomposition} realizes the dissipaton decomposition of the bath force used in the literature \cite{Yan14054105,Wan22170901}, here as an explicit decomposition of its left and right actions in Liouville space. Equation~\eqref{eq:dissipaton_algebra} is the corresponding generalized Wick's theorem: an inserted dissipaton either raises the occupation by one or contracts with one of the $n_k$ existing factors, lowering it by one. The left and right contractions carry $\eta_k$ and $\eta_{\bar k}^*$, respectively. Summing these insertions in the interaction Liouvillian reproduces the upper- and lower-tier terms of \Eq{single-normal-heom}. Thus, the dissipaton decomposition and generalized Wick's theorem follow here from explicit bath operators and their duals, rather than being imposed as separate algebraic prescriptions. 

\subsection{Concluding remarks}

This section constructs the continuous-bath hierarchy directly from ordered bath moments. For the nonnegative rational spectra admitting the stated stable spectral factor, the thermofield representation yields explicit dissipaton operators as smeared annihilation fields. Their nonorthogonal $\mathbf M$ matrix determines the equal-time contractions, and its row sums recover the generally complex amplitudes of the exponential correlation expansion. The causal filter representation supplies the Langevin evolution with a common vacuum input. For the factorized vacuum preparation, causality and normal ordering then give the generalized diffusion relation through a fixed-state forward derivative, leading to the double-index hierarchy.

The dual operators resolve the cross-mode contractions into Liouville-space insertion maps while preserving the total force action. Pairing conjugate decay rates further projects the double-index moments onto a closed single-index hierarchy. In this representation, the dissipaton decomposition and generalized Wick's theorem \cite{Yan14054105} take the forms of \Eqs{single-force-decomposition} and (\ref{eq:dissipaton_algebra}), providing an explicit operator realization of the established dissipaton algebra. The construction separates the microscopic, nonorthogonal bath modes from the mode-resolved insertion superoperators and connects both to the same reduced dynamics. The normal-ordered hierarchy serves as the reference for higher powers of the bath force in Sec.~\ref{sec:quadratic} and for other orderings in Appendix~\ref{app:other-orderings}.

The ODO incorporates the dynamics of system--bath correlations, thereby retaining information beyond that contained in the reduced system density operator alone. Its evolution reflects not only the dynamics of the system itself but also the development of correlations between the system and its environment. This feature is essential for characterizing the coupled system--bath dynamics, as discussed in detail in Appendix~\ref{app:hermite}.

\section{Nonlinear system--bath coupling}\label{sec:quadratic}

In the previous section, we constructed the dissipaton operators and derived the corresponding normal-ordered hierarchy for linear system--bath coupling, recovering the influence-functional result. We now consider polynomial system--bath interactions of the form
\begin{align}\label{nonlinear-coupling}
 H_{\SB} = \sum_{r=1}^{R} \hat Q_r \hat F^r,
\end{align}
where $\hat Q_r=\hat Q_r^\dagger$ are system operators that include the coupling coefficients. We assume a well-defined total evolution and finite moments for the observables considered. For a Gaussian bath nonlinearly coupled to the system, the influence functional remains fully determined by the bare-bath correlation function $C(t)=\la\hat F_{\B}(t)\hat F_{\B}(0)\ra_{\B}$. Due to nonlinear terms, however, the cumulant expansion generally contains infinitely many nonzero terms, so the simple second-order cumulant expression available for linear coupling no longer applies. The dissipaton construction remains applicable because it relies on the free-bath generalized Langevin equation and the dissipaton commutation relations, neither of which depends on the form of the system--bath interaction. We first treat quadratic coupling and then extend the formulation to polynomial coupling of arbitrary finite order.

\subsection{Quadratic case}

We directly derive the single-index hierarchy, that is, starting with the ODOs defined in \Eq{single-index-odo}. In the bath interaction picture, we have 
\begin{align}
  \rho_{\bm n}(t) = {\rm tr}_{\B}\bigg[ \mathcal N\bigg(\prod_k[\hat\phi_k^{\B}(t)]^{n_k}\bigg)\rho_{\T}^{I}(t) \bigg],
\end{align}
with 
\begin{align}
  \dot\rho_{\T}^I(t) = -i[H_{\tS} + \hat Q_1\hat F_{\B}(t) + \hat Q_2\hat F_{\B}^2(t), \rho_{\T}^I(t)].
\end{align}
The derivative of the dissipaton moments is resolved by the generalized diffusion equation, \Eq{continuous-normal-diffusion}. For the quadratic term, we first decompose the bath force as in \Eq{single-force-decomposition} and then apply the generalized Wick's theorem, \Eq{eq:dissipaton_algebra}. Then we have 
\begin{align}
 {\rm tr}_{\B}[\mathcal N(\bm n)\hat F^{>}\hat F^{>}\rho_{\T}] &=  \sum_{kl}{\rm tr}_{\B}[\mathcal N(\bm n)\mathscr f_k^>\mathscr f_l^>\rho_{\T}] \nl
  &= \sum_{kl}\rho_{{\bm n}_{kl}^{++}} + \la\hat F_{\B}^2\ra_{\B}\rho_{\bm n}  \nl
  &\quad\, + 2\sum_{kl}n_k\eta_k\rho_{{\bm n}_{kl}^{-+}} \nl
  &\quad\, + \sum_{kl}n_k(n_l-\delta_{kl})\eta_k\eta_l\rho_{{\bm n}_{kl}^{--}}.
\end{align}
The expression for ${\rm tr}_{\B}[\mathcal N(\bm n)\hat F^{<}\hat F^{<}\rho_{\T}]$ is similar, with $\eta_k$ replaced by $\eta_{\bar k}^*$. Consequently, the single-index hierarchy for quadratic coupling is
\begin{align}\label{single-quadratic-heom}
  \dot\rho_{\bm n} &= -i[H_{\tS},\rho_{\bm n}] - \sum_k n_k\gamma_k\rho_{\bm n} -i \la\hat F^2\ra_{\B}[\hat Q_2,\rho_{\bm n}] \nl
  &\quad\, -i\sum_k[\hat Q_1,\rho_{\bm n_k^+}] -i \sum_{kk'}[\hat Q_2,\rho_{\bm n_{kk'}^{++}}] \nl
  &\quad\, -i\sum_k n_k(\eta_k\hat Q_1\rho_{\bm n_k^-}-\eta_{\bar k}^*\rho_{\bm n_k^-}\hat Q_1) \nl
  &\quad\, -2i\sum_{kl}n_k(\eta_k\hat Q_2\rho_{{\bm n}_{kl}^{-+}}-\eta_{\bar k}^*\rho_{{\bm n}_{kl}^{-+}}\hat Q_2) \nl
  &\quad\, -i\sum_{kl}n_k(n_l - \delta_{kl})\eta_k\eta_l\hat Q_2\rho_{{\bm n}_{kl}^{--}} \nl
  &\quad\, +i\sum_{kl}n_k(n_l - \delta_{kl})\eta_{\bar k}^*\eta_{\bar l}^* \rho_{{\bm n}_{kl}^{--}}\hat Q_2.
\end{align}
This is the extended DEOM for quadratic coupling \cite{Xu17395,Xu18114103}. 

\subsection{General polynomial case}\label{sec:polynomial}

The generalized Wick's theorem, \Eq{eq:dissipaton_algebra}, is a recursion for force insertions. It can therefore generate the hierarchy for any finite polynomial interaction without deriving a separate contraction formula for each power. Define 
\begin{align}\label{polynomial-insertion-moments}
 \hat X_{\bm n}^{(r)\gtrless}
 &\equiv {\rm tr}_{\B}\left[\mathcal N(\bm n)
 (\hat F^{\gtrless})^r\rho_{\T}\right],
\end{align}
and write $\eta_k^>\equiv\eta_k$ and $\eta_k^<\equiv\eta_{\bar k}^*$. Because the insertion identities hold for any operator in place of $\rho_{\T}$, they give
\begin{equation}\label{polynomial-recursion}
 \hat X_{\bm n}^{(r+1)\gtrless}
 =\sum_k\left[\hat X_{\bm n_k^+}^{(r)\gtrless}
       +n_k\eta_k^{\gtrless}\hat X_{\bm n_k^-}^{(r)\gtrless}\right].
\end{equation}
Each step either adds a normally ordered factor or contracts one existing factor. In particular, applying the recursion twice reproduces the quadratic expressions above, including the constant contraction $C(t=0)=\sum_k\eta_k=\la\hat F^2\ra_{\B}$.

For \Eq{nonlinear-coupling}, the resulting single-index hierarchy is
\begin{align}\label{polynomial-heom}
 \dot\rho_{\bm n}={}&-i[H_{\tS},\rho_{\bm n}]
 -\sum_k n_k\gamma_k\rho_{\bm n}\nl
 &-i\sum_{r=1}^{R}\left[
 \hat Q_r \hat X_{\bm n}^{(r)>}
 -\hat X_{\bm n}^{(r)<}\hat Q_r\right].
\end{align}
The free diffusion term remains unchanged. Indeed, the future input increments in \Eq{continuous-future-commutator} commute with every polynomial in $\hat F_{\B}(\tau)$ for $\tau\leq t$. Thus the causality argument leading to \Eq{continuous-normal-diffusion} also applies to the nonlinear propagator, without assuming that the interacting bath state remains Gaussian.

Equations~\eqref{polynomial-recursion} and \eqref{polynomial-heom} provide a direct algorithm for symbolic equation generation. For each target index $\bm n$, start from $\hat X_{\bm n}^{(0)\gtrless}=\rho_{\bm n}$, apply the two insertion rules $r$ times, and collect terms with identical final indices. An implementation can store each intermediate expression as a sparse map from occupation vectors to scalar coefficients; repeated intermediate expressions can be reused. The occupation factors in each step are evaluated at that step's indices, so repeated-index contractions and their combinatorial multiplicities are generated automatically. A term of degree $r$ connects only tiers whose differences from $|\bm n|\equiv\sum_kn_k$ belong to $\{-r,-r+2,\cdots,r\}$. The algebraic generation is finite for any specified $r$ and target index, although dynamical propagation still requires a hierarchy truncation and convergence checks. Truncation should be imposed after the insertion polynomial is assembled, since an intermediate upward step may subsequently contract back into a retained tier.


\section{Discussion and conclusions}\label{sec:conclusions}
We have constructed a statistical quasiparticle representation that connects explicit bath operators to nonlinear hierarchical dynamics. For a nonnegative rational spectrum admitting the stated stable factor, the dissipatons are smeared thermofield annihilation operators with specified kernels and adjoints. Their nonorthogonal Gram matrix determines the Wick contractions, and its row sums give the complex correlation amplitudes. A common vacuum input supplies the Langevin evolution and the fixed-state forward diffusion relation. Dual operators then realize the mode-resolved dissipaton insertions in Liouville space. These ingredients yield polynomial force couplings through the finite insertion recursion of Sec.~\ref{sec:polynomial}, with the quadratic case recovering the extended DEOM. The recursion specifies a practical route to automatic equation generation, including repeated-index contractions, without a new analytical derivation for each interaction order.

The construction is closely related to auxiliary-mode representations. Pseudomodes already admit explicit oscillator descriptions, and Fano mappings provide bath-mode realizations in suitable models \cite{Dal01053813,Ple20043058}; nonlinear pseudomode formulations are also available \cite{Zha25115131}. Here the exponential kernels are retained as nonorthogonal modes so that their simple drift, cross contractions, and dissipaton insertion algebra remain visible in the moment hierarchy. Appendix~\ref{app:langevin} makes the connection to coupled Lindblad modes explicit. The contribution is the resulting operator derivation of the dissipaton moment theory and its nonlinear force rules, rather than the existence of an auxiliary-mode representation alone.

The same operator framework also clarifies the meaning of the hierarchy variables. The continuous-bath $G$ transformation relates normal, Weyl, and anti-normal moments through the full nonorthogonal contraction matrix; transforming the diffusion terms and initial conditions is essential to their equivalence. The generalized Hermite construction distinguishes mode-resolved Liouville-space contractions from physical cross-mode contractions and converts the ODOs into bath-force moments, collective-mode observables, and system--bath correlations. These reconstruction identities apply to the interacting state, even when nonlinear coupling makes it non-Gaussian.


\begin{acknowledgments}
Support from the National Natural Science Foundation of China (Grant No.~224B2305) is gratefully acknowledged. The authors are indebted to YiJing Yan and Bing Gu for invaluable discussions. ChatGPT (OpenAI) was used to improve the clarity and language of the manuscript.
\end{acknowledgments}

\appendix

\section{Different orderings and Gaussian \texorpdfstring{$G$}{G} transformations}\label{app:other-orderings}

Section~\ref{sec:continuous} established the normal-ordered dissipaton hierarchy and its single-index projection. We now use them to obtain the Weyl- and anti-normal-ordered representations and the transformations relating all three. Changing the ordering modifies the auxiliary moments, their initial values, and the contraction terms in their dynamics. The Gaussian $G$ transformation organizes these changes using the same nonorthogonal commutator matrix $\bf M$; no new bath representation is required.

\subsection{Ordered moments and the Gaussian \texorpdfstring{$G$}{G} transformation}\label{sec:continuous-G}
The normal hierarchy of \Sec{sec:continuous} provides the reference. A common ordering parameter $s$ is used for all modes, with $\kappa\equiv(1-s)/2$. Normal, Weyl, and anti-normal orderings correspond to $(s,\kappa)=(1,0),(0,1/2),(-1,1)$, respectively. For the nonorthogonal dissipatons, the global ordered products are defined through
\begin{align}\label{continuous-ordering-generating}
 \sum_{\bm u,\bm v}\prod_{k=1}^K
 \frac{z_k^{u_k}w_k^{v_k}}{u_k!v_k!}
 \mathcal O_s(\bm u,\bm v)=e^{\kappa\bm z^T\mathbf M\bm w}
 e^{\sum_jw_j\hat f_j^\dagger}e^{\sum_i z_i\hat f_i}.
\end{align}
Using the Baker--Campbell--Hausdorff identity, with
$[\sum_i z_i\hat f_i,\sum_jw_j\hat f_j^\dagger]=\bm z^T\mathbf M\bm w$,
identifies $s=0$ with Weyl ordering and $s=-1$ with anti-normal ordering. In particular, the generating operator at $s=0$ is $e^{\sum_i(z_i\hat f_i+w_i\hat f_i^\dagger)}$, while at $s=-1$ it is $e^{\sum_i z_i\hat f_i}e^{\sum_jw_j\hat f_j^\dagger}$. Thus, the change of ordering contracts different modes through $M_{ij}$ as well as contracting each mode with itself.

Define ordered density operators in arbitrary ordering $s$ as
\begin{align}
  \rho^{(s)}_{\bm u,\bm v}\equiv {\rm tr}_{\B}[\mathcal O_s(\bm u,\bm v)\rho_{\T}]
\end{align}
and its generating function,
\begin{equation}\label{genfun}
 \mathcal Z_s(\bm z,\bm w)
 \equiv\sum_{\bm u,\bm v}\prod_{k=1}^{K}\frac{z_k^{u_k}w_k^{v_k}}{u_k!v_k!}\rho^{(s)}_{\bm u,\bm v}.
\end{equation}
Here $\bm z,\bm w$ are independent formal variables, and $\rho^{(\mathcal N)}_{\bm u,\bm v}=\rho_{\bm u,\bm v}$ denotes the normal ODOs of the main text. Multiplying $\rho_{\T}$ and tracing over bath of \Eq{continuous-ordering-generating} gives
\begin{equation}\label{continuous-G-generating}
 \mathcal Z_s=e^{\kappa\bm z^T\mathbf M\bm w}\mathcal Z_{\mathcal N}
 \equiv e^{-\frac12\bm x^T\mathbf G_s\bm x}\mathcal Z_{\mathcal N},
\end{equation}
where
\begin{equation}\label{continuous-G-matrix}
 \bm x=\begin{pmatrix}\bm z\\\bm w\end{pmatrix},
 \quad \mathbf G_s=-\kappa\begin{pmatrix}0& \mathbf M\\  \mathbf M^T&0\end{pmatrix}.
\end{equation}
The transpose in the lower block follows from the bilinear form in the independent formal variables $\bm z,\bm w$. 

To proceed, we define the collection $\bm\rho^{(s)} \equiv \{\rho_{\bf u,v}^{(s)}\}$ and the hierarchy index maps by
\begin{equation}\label{index-ops}
 \begin{split}
  (\bm{\mathcal R}_{u_k}\boldsymbol\rho)_{\bm u,\bm v}
 &\equiv \rho_{\bm u_k^+,\bm v},\quad
 (\bm{\mathcal D}_{u_k}\boldsymbol\rho)_{\bm u,\bm v}
 \equiv u_k\rho_{\bm u_k^-,\bm v},\\
 (\bm{\mathcal R}_{v_k}\boldsymbol\rho)_{\bm u,\bm v}
 &\equiv \rho_{\bm u,\bm v_k^+},\quad
 (\bm{\mathcal D}_{v_k}\boldsymbol\rho)_{\bm u,\bm v}
 \equiv v_k\rho_{\bm u,\bm v_k^-}.
 \end{split}
\end{equation}
They satisfy $[\bm{\mathcal R}_{u_k},\bm{\mathcal D}_{u_l}]=\delta_{kl}\bm{\mathcal I}$ and $[\bm{\mathcal R}_{v_k},\bm{\mathcal D}_{v_l}]=\delta_{kl}\bm{\mathcal I}$, where $\bm{\mathcal I}$ is the hierarchy identity; all other commutators vanish.  Let $\bm{\mathcal E}$ denote the map taking any hierarchy $\boldsymbol\rho$ to its exponential generating function as in \Eq{genfun}. A direct index shift gives
\begin{align}\label{app:dictionary}
\begin{split}
     \bm{\mathcal E}(\bm{\mathcal D}_{u_k}\boldsymbol\rho)
 &=z_k\bm{\mathcal E}(\boldsymbol\rho),\quad
 \bm{\mathcal E}(\bm{\mathcal D}_{v_k}\boldsymbol\rho)
 =w_k\bm{\mathcal E}(\boldsymbol\rho),\\
 \bm{\mathcal E}(\bm{\mathcal R}_{u_k}\boldsymbol\rho)
 &=\partial_{z_k}\bm{\mathcal E}(\boldsymbol\rho),\quad
 \bm{\mathcal E}(\bm{\mathcal R}_{v_k}\boldsymbol\rho)
 =\partial_{w_k}\bm{\mathcal E}(\boldsymbol\rho).
\end{split}
\end{align}
Namely, $\bm{\mathcal D}$ is represented by multiplication, whereas $\bm{\mathcal R}$ is represented by differentiation. As a consequence, for 
\begin{align}
  \bm{\mathcal C}_M\equiv\sum_{kl}M_{kl}\bm{\mathcal D}_{u_k}\bm{\mathcal D}_{v_l},
\end{align}
we have
\begin{equation}\label{app:intertwining}
 \bm{\mathcal E}(\bm{\mathcal C}_M\boldsymbol\rho)
 =(\bm z^T\mathbf M\bm w)\bm{\mathcal E}(\boldsymbol\rho).
\end{equation}
Applying this identity to every term of the exponential yields
\begin{equation}\label{app:exp-map}
 \bm{\mathcal E}(e^{\kappa\bm{\mathcal C}_M}\boldsymbol\rho)
 =e^{\kappa\bm z^T\mathbf M\bm w}\bm{\mathcal E}(\boldsymbol\rho).
\end{equation}
Since $\bm{\mathcal E}$ is injective by coefficient extraction, \Eq{app:exp-map} is equivalent to 
\begin{equation}\label{continuous-G-hierarchy}
 \boldsymbol\rho^{(s)}=\bm{\mathcal V}_s\boldsymbol\rho^{(\mathcal N)}
\end{equation}
with
\begin{equation}\label{continuous-contraction-map}
\bm{\mathcal V}_s=e^{\kappa\bm{\mathcal C}_M}.
\end{equation}
The inverse is $\bm{\mathcal V}_s^{-1}=e^{-\kappa\bm{\mathcal C}_M}$, and the transformation between any two common orderings is $\bm{\mathcal V}_{s\leftarrow s'}=e^{(s'-s)\bm{\mathcal C}_M/2}$.

The map contracts one annihilation and one creation factor at a time, including factors from different modes because $\mathbf M$ is nondiagonal. For any fixed output tier, its exponential contains only finitely many nonzero terms. It changes the definition of the auxiliary moments while leaving their zeroth component, the reduced density operator, unchanged. The same transformation must be applied to both the generator and the initial moments to compare different orderings.

\subsection{Transformed dynamics and free diffusion}\label{sec:ordered-diffusion}
To transform the dynamics, first collect the normal ODOs into $\boldsymbol\rho^{(\mathcal N)}$. Using the bold index superoperators defined in \Eq{index-ops}, we can write $\dot{\boldsymbol\rho}^{(\mathcal N)}=-i\bm{\mathcal L}_\mathcal N\boldsymbol\rho^{(\mathcal N)}$, where 
\begin{align}\label{continuous-hierarchy-generator}
 -i\bm{\mathcal L}_\mathcal N\equiv {}&-i\mathcal L_{\tS}
 -\sum_k\left(\gamma_k\bm{\mathcal M}_{u_k}
             +\gamma_k^*\bm{\mathcal M}_{v_k}\right)\nl
 &-i\hat Q^\times\sum_k
 \left(\bm{\mathcal R}_{u_k}+\bm{\mathcal R}_{v_k}\right)\nl
 &-i\sum_k\left(\eta_k\hat Q^>\bm{\mathcal D}_{u_k}
             -\eta_k^*\hat Q^<\bm{\mathcal D}_{v_k}\right).
\end{align}
Here, $\mathcal L_{\tS}\equiv H_{\tS}^\times$, $(\bm{\mathcal M}_{u_k}\boldsymbol\rho)_{\bm u,\bm v}=u_k\rho_{\bm u,\bm v}$ and $(\bm{\mathcal M}_{v_k}\boldsymbol\rho)_{\bm u,\bm v}=v_k\rho_{\bm u,\bm v}$ count the two occupations.

The equations of motion transform by the similarity relation
\begin{equation}\label{continuous-G-similarity}
 \bm{\mathcal L}_s=\bm{\mathcal V}_s
 \bm{\mathcal L}_{\mathcal N}\bm{\mathcal V}_s^{-1},
 \quad\dot{\boldsymbol\rho}^{(s)}=-i\bm{\mathcal L}_s\boldsymbol\rho^{(s)}.
\end{equation}
The index commutation relations give
\begin{subequations}\label{continuous-G-index-action}
    \begin{align}
 \bm{\mathcal V}_s\bm{\mathcal R}_{u_i}\bm{\mathcal V}_s^{-1}
 &=\bm{\mathcal R}_{u_i}-\kappa\sum_jM_{ij}\bm{\mathcal D}_{v_j},\\
 \bm{\mathcal V}_s\bm{\mathcal R}_{v_j}\bm{\mathcal V}_s^{-1}
 &=\bm{\mathcal R}_{v_j}-\kappa\sum_iM_{ij}\bm{\mathcal D}_{u_i},
\end{align}
\end{subequations}
while the $\bm{\mathcal D}$ operators remain unchanged. The damped mode contribution satisfies
\begin{equation}\label{continuous-G-free-commutator}
 \left[\bm{\mathcal C}_M,
 \sum_k(\gamma_k\bm{\mathcal M}_{u_k}
       +\gamma_k^*\bm{\mathcal M}_{v_k})\right]
 =-2\sum_{ij}r_i r_j^*\bm{\mathcal D}_{u_i}\bm{\mathcal D}_{v_j}.
\end{equation}
Here we have used $(\gamma_i+\gamma_j^*)M_{ij}=2r_i r_j^*$. Consequently, with the same fixed-state right-derivative definition as in \Eq{continuous-normal-diffusion}, the free diffusion relation for an arbitrary common ordering is
\begin{align}\label{continuous-s-diffusion}
 \left.\partial_t^+\rho^{(s)}_{\bm u,\bm v}\right|_{\B}
 ={}&-\sum_k(u_k\gamma_k+v_k\gamma_k^*)
 \rho^{(s)}_{\bm u,\bm v}\nl
 &+2\kappa\sum_{ij}u_i v_j r_i r_j^*
 \rho^{(s)}_{\bm u_i^-,\bm v_j^-}.
\end{align}
The second term lowers the total tier by two. It vanishes for normal ordering, but is required in Weyl and anti-normal ordering because their moments retain vacuum contractions. The dissipative drift and the change of ordering do not commute: the identity $(\gamma_i+\gamma_j^*)M_{ij}=2r_i r_j^*$ fixes the additional diffusion term.

The origin of this term can also be seen directly from the finite-time bath evolution. For the vacuum increment $\hat\nu_{k}(t;\Delta t)$ introduced in \Sec{sec:continuous-hierarchy}, direct integration gives
\begin{align}\label{continuous-wavepacket-covariance}
 [\hat\nu_{i}(t;\Delta t),\hat\nu_{j}(t;\Delta t)^\dagger]
 &=2r_i r_j^*\int_t^{t+\Delta t}\ud \tau\,
 e^{-(\gamma_i+\gamma_j^*)(t+\Delta t-\tau)}\nl
 &=M_{ij}\bigl[1-e^{-(\gamma_i+\gamma_j^*)\Delta t}\bigr].
\end{align}
For any operator $\hat X_t$ commuting with these operators, \Eq{continuous-future-vacuum} consequently gives
\begin{subequations}\label{continuous-wavepacket-traces}
\begin{align}
 {\rm tr}_{\B}[\hat X_t\hat\nu_{j}(t;\Delta t)^\dagger\hat\nu_{i}(t;\Delta t)\rho_{\T}^{I}(t)]&=0,\\
 {\rm tr}_{\B}[\hat X_t\hat\nu_{i}(t;\Delta t)\hat\nu_{j}(t;\Delta t)^\dagger\rho_{\T}^{I}(t)]&= \nl
 &\hspace{-70pt}M_{ij}(1-e^{-\lambda_{ij}\Delta t}){\rm tr}_{\B}[\hat X_t\rho_{\T}^{I}(t)].
\end{align}
\end{subequations}
where $\lambda_{ij}=\gamma_i+\gamma_j^*$. These finite-interval bath-trace identities supply an operator basis for the diffusion relation. 
For a second-order example, define $\hat P_{ij}^{(s)\B}(t)\equiv\hat f_j^{\dagger\B}(t)\hat f_i^{\B}(t)+\kappa M_{ij}$. Equation~(\ref{continuous-normal-finite-trace}) implies
\begin{align}\label{continuous-ordering-pair-finite}
 {\rm tr}_{\B}[\hat P_{ij}^{(s)\B}(t+\Delta t)\rho_{\T}^{I}(t)]
 ={}&e^{-\lambda_{ij}\Delta t}{\rm tr}_{\B}[\hat P_{ij}^{(s)\B}(t)\rho_{\T}^{I}(t)]\nl
 &+\kappa M_{ij}(1-e^{-\lambda_{ij}\Delta t})\rho_{\tS}(t).
\end{align}
Its right derivative is
\begin{align}\label{continuous-ordering-pair-drift}
 &\quad\,\lim_{\Delta t\to0^+}\frac{1}{\Delta t}{\rm tr}_{\B}\bigl[
 (\hat P_{ij}^{(s)\B}(t+\Delta t)-\hat P_{ij}^{(s)\B}(t))\rho_{\T}^{I}(t)\bigr]\nl
 &=-\lambda_{ij}{\rm tr}_{\B}[\hat P_{ij}^{(s)\B}(t)\rho_{\T}^{I}(t)]
 +2\kappa r_i r_j^*\rho_{\tS}(t).
\end{align}
This is the second-order instance of \Eq{continuous-s-diffusion}; it shows explicitly why the diffusion term is needed even for a factorized vacuum state.

\subsection{Normal, Weyl, and anti-normal hierarchies}\label{sec:ordered-results}
Combining \Eqs{continuous-hierarchy-generator} and (\ref{continuous-G-index-action})--(\ref{continuous-G-free-commutator}), we obtain
\begin{align}\label{continuous-s-heom}
 \dot\rho^{(s)}_{\bm u,\bm v}
 ={}&-\left[i\mathcal L_{\tS}
 +\sum_k(u_k\gamma_k+v_k\gamma_k^*)\right]
 \rho^{(s)}_{\bm u,\bm v}\nl
 &-i\hat Q^\times\sum_k\left(
 \rho^{(s)}_{\bm u_k^+,\bm v}
 +\rho^{(s)}_{\bm u,\bm v_k^+}\right)\nl
 &-i\sum_k u_k\eta_k
 [(1-\kappa)\hat Q^>+\kappa\hat Q^<]
 \rho^{(s)}_{\bm u_k^-,\bm v}\nl
 &+i\sum_k v_k\eta_k^*
 [\kappa\hat Q^>+(1-\kappa)\hat Q^<]
 \rho^{(s)}_{\bm u,\bm v_k^-}\nl
 &+2\kappa\sum_{ij}u_i v_j r_i r_j^*
 \rho^{(s)}_{\bm u_i^-,\bm v_j^-}.
\end{align}
The three orderings follow by choosing $\kappa$. Below, 
\begin{align}
  \hat Q^\Diamond \equiv \hat Q^>+\hat Q^<
\end{align}
denotes the anticommutator action.

\emph{Normal ordering} ($s=1$, $\kappa=0$) recovers \Eq{continuous-normal-heom}. The downward interaction terms contain $\hat Q^>$ for the annihilation indices and $\hat Q^<$ for the creation indices, and no two-tier diffusion term remains.

\emph{Weyl ordering} ($s=0$, $\kappa=1/2$) assigns equal weight to the two multiplication orders. The downward interaction terms are
\begin{align*}
 &-\frac{i}{2}\sum_k u_k\eta_k\hat Q^\diamond
 \rho^{(0)}_{\bm u_k^-,\bm v}
 +\frac{i}{2}\sum_k v_k\eta_k^*\hat Q^\diamond
 \rho^{(0)}_{\bm u,\bm v_k^-},
\end{align*}
while the diffusion coefficient $2\kappa r_i r_j^*$ becomes $r_i r_j^*$.

\emph{Anti-normal ordering} ($s=-1$, $\kappa=1$) exchanges the left and right system actions in the downward terms relative to normal ordering:
\begin{align*}
 &-i\sum_k u_k\eta_k\hat Q^<
 \rho^{(-1)}_{\bm u_k^-,\bm v}
 +i\sum_k v_k\eta_k^*\hat Q^>
 \rho^{(-1)}_{\bm u,\bm v_k^-}.
\end{align*}
The diffusion coefficient is $2r_i r_j^*$. In all three cases, the diagonal drift and the upward interaction terms have the same form; the ordering determines the downward contractions and the two-tier diffusion contribution.
 
For the factorized preparation, the initial generating function is
\begin{equation}\label{continuous-s-initial}
 \mathcal Z_s(\bm z,\bm w;0)
 =e^{\kappa\bm z^TM\bm w}\rho_{\tS}(0).
\end{equation}
In particular, $\rho^{(s)}_{\mathbf 0^+_i,\mathbf 0^+_j}(0)=\kappa M_{ij}\rho_{\tS}(0)$, including cross-mode initial contractions. These initial auxiliary moments encode the chosen ordering, not initial system--bath correlations. Since the zeroth component of $\bm{\mathcal V}_s\boldsymbol\rho$ equals $\rho_{\bm0,\bm0}$, the full hierarchies yield the same reduced dynamics when initialized consistently. This equivalence applies before hierarchy truncation; convergence of a finite-tier calculation must be checked for the chosen representation.

\subsection{Single-index hierarchies and their \texorpdfstring{$G$}{G} transformation}\label{sec:single-index}
The conjugate-mode projection of \Sec{sec:continuous} also applies to the ordered hierarchy. We retain $\hat\phi_k=\hat f_k+\hat f_{\bar k}^\dagger$ and introduce the symmetric contraction matrix
\begin{align}\label{single-covariance}
 B_{ij}&\equiv M_{i\bar j}+M_{j\bar i}=B_{ji},\\
 W_{ij}&\equiv(\gamma_i+\gamma_j)B_{ij}
 =2(r_i r_{\bar j}^*+r_j r_{\bar i}^*).
\end{align}
The single-index ODOs are defined by
\begin{equation}\label{single-projection}
 \rho^{(s)}_{\bm n}
 =\sum_{\bm m\leq\bm n}\binom{\bm n}{\bm m}
 \rho^{(s)}_{\bm m,\overline{\bm n-\bm m}},
\end{equation}
with $(\overline{\bm m})_k\equiv m_{\bar k}$.
Putting $z_k=x_k$ and $w_k=x_{\bar k}$ in \Eq{continuous-G-generating} gives 
\begin{align}
  \mathcal Y_s=e^{\kappa\bm x^T\mathbf B\bm x/2}\mathcal Y_{\mathcal N},
\end{align}
where
\begin{align}
  \mathcal Y_s=\sum_{\bm n}\prod_{k=1}^{K}\frac{x_k^{n_k}}{n_k!}\rho^{(s)}_{\bm n}.
\end{align}
Thus projecting and changing the ordering give the same result. The single-index $G$ transformation is
\begin{equation}\label{single-G}
 \boldsymbol\sigma^{(s)}=\bm{\mathcal V}^{\phi}_s
 \boldsymbol\sigma^{(\mathcal N)},\quad
 \bm{\mathcal V}^{\phi}_s
 =\exp\left(\frac\kappa2\sum_{ij}B_{ij}
 \bm{\mathcal D}_i\bm{\mathcal D}_j\right),
\end{equation}
where $\boldsymbol\sigma\equiv \{\rho_{\bm n}\}$ and $(\bm{\mathcal D}_i\boldsymbol\sigma)_{\bm n}\equiv n_i\rho_{\bm n_i^-}$. We also define $(\bm{\mathcal R}_i\boldsymbol\sigma)_{\bm n}=\rho_{\bm n_i^+}$. Between two arbitrary common orderings, the transformation is $\bm{\mathcal V}^{\phi}_{s\leftarrow s'}=\exp[(s'-s)\sum_{ij}B_{ij}\bm{\mathcal D}_i\bm{\mathcal D}_j/4]$. 

Applying \Eq{single-G} to \Eq{single-normal-heom} gives
\begin{align}\label{single-s-heom}
 \dot\rho^{(s)}_{\bm n}
 ={}&-\left(i\mathcal L_{\tS}+\sum_k n_k\gamma_k\right)
 \rho^{(s)}_{\bm n}
 -i\hat Q^\times\sum_k\rho^{(s)}_{\bm n_k^+}\nl
 &-i\sum_k n_k
 \bigl\{[(1-\kappa)\eta_k-\kappa\eta_{\bar k}^*]\hat Q^>\nl
 &\hspace{43pt}-[(1-\kappa)\eta_{\bar k}^*-\kappa\eta_k]\hat Q^<\bigr\}
 \rho^{(s)}_{\bm n_k^-}\nl
 &+\frac\kappa2\sum_{ij}W_{ij}n_i(n_j-\delta_{ij})
 \rho^{(s)}_{\bm n_{ij}^{--}}.
\end{align}
Here, the factor $n_i(n_j-\delta_{ij})$ counts two distinct factors, including when they belong to the same mode. The interaction coefficients use $\sum_jB_{ij}=\eta_i+\eta_{\bar i}^*$, while $W_{ij}$ determines the projected diffusion.

Normal, Weyl, and anti-normal hierarchies again follow by taking $\kappa=0,1/2,1$, respectively. The factorized thermal preparation requires
\begin{equation}\label{single-initial}
 \mathcal Y_s(\bm x;0)
 =e^{\kappa\bm x^T\mathbf B\bm x/2}\rho_{\tS}(0).
\end{equation}
Consequently, $\rho^{(s)}_{\mathbf 0_{ij}^{++}}(0)
 =\kappa B_{ij}\rho_{\tS}(0)$, with higher even moments given by pair contractions and odd moments equal to zero. Only the normal hierarchy starts with all auxiliary ODOs vanishing.

The matrices $\bf M$ and $\bf B$ therefore specify the same Gaussian change of ordering before and after projection. Normal, Weyl, and anti-normal moments are alternative representations of the same bath dynamics, with the transformed initial conditions accounting for their different vacuum contractions. The same transformed force insertion maps can be iterated for the polynomial interactions of Sec.~\ref{sec:quadratic}.

\section{Generalized Hermite polynomials and environmental observables}\label{app:hermite}

The ODOs encode ordered environmental moments rather than ordinary powers of independent bath coordinates. The insertion maps in \Eq{single-insertion-maps} and their generalized Wick recursion, \Eq{eq:dissipaton_algebra}, specify the polynomial conversion between these quantities. We first formulate this conversion in Liouville space, where the contractions are diagonal in the mode labels, and then relate it to physical bath operators.


For commuting variables $\bm y=(y_1,\ldots,y_K)$ and coefficients $\bm c=(c_1,\ldots,c_K)$, define
\begin{align}\label{app:hermite-definition}
 H_{\bm n}(\bm y;\bm c)
 \equiv\left.\prod_k\frac{\partial^{n_k}}{\partial x_k^{n_k}}
 \exp\left(\sum_kx_ky_k-\tfrac12\sum_kc_kx_k^2\right)
 \right|_{\bm x=\bm0}.
\end{align}
These generalized Hermite polynomials are well defined also for complex $c_k$; no positive probability measure is assumed. They obey
\begin{equation}\label{app:hermite-recursion}
 H_{\bm n_k^+}=y_kH_{\bm n}-n_kc_kH_{\bm n_k^-},\quad H_{\bm0}=1.
\end{equation}
These lead to the expansion
\begin{equation}\label{app:hermite-expansion}
 H_{\bm n}(\bm y;\bm c)
 =\sum_{2\bm m\leq\bm n}
 \prod_k\frac{n_k!(-c_k/2)^{m_k}y_k^{n_k-2m_k}}
 {m_k!(n_k-2m_k)!}.
\end{equation}

Within each insertion branch, the superoperators $\{\mathscr f_k^{\gtrless}\}$ commute. We may therefore substitute $y_k=\mathscr f_k^{\gtrless}$ and $c_k=\eta_k^{\gtrless}$, with $\eta_k^>=\eta_k$ and $\eta_k^<=\eta_{\bar k}^*$. Define $\mathscr H_{\bm n}^{\gtrless}(\bm{\mathscr f}^{\gtrless})\equiv H_{\bm n}(\bm{\mathscr f}^{\gtrless};\bm c^{\gtrless})$. Then we have \cite{Fan112145}
\begin{equation}\label{app:hermite-odo}
 \rho_{\bm n}(t)
 ={\rm tr}_{\B}\left[\mathscr H_{\bm n}^>(\bm{\mathscr f}^{\gtrless})\rho_{\T}(t)\right]
 ={\rm tr}_{\B}\left[\mathscr H_{\bm n}^<(\bm{\mathscr f}^{\gtrless})\rho_{\T}(t)\right].
\end{equation}
This is an identity of moment maps and remains valid for the correlated, generally non-Gaussian total state produced by nonlinear coupling. The diagonal contractions in \Eq{app:hermite-definition} belong to the Liouville-space insertion maps. They must not be identified with $\la\hat\phi_i\hat\phi_j\ra_{\B}=M_{i\bar j}$, which is generally nondiagonal. In particular, replacing the Hilbert-space operators $\hat\phi_k$ for $\mathscr f_k^{\gtrless}$ in \Eq{app:hermite-odo} would generally be incorrect.

Equation~\eqref{app:hermite-odo} has an inverse relation. From \Eq{app:hermite-definition}, we have 
\begin{align}
  e^{\sum_k x_ky_k} = e^{\sum_kc_kx_k^2/2}\sum_{\bm l}\prod_k\frac{x_k^{l_k}}{l_k!}H_{\bm l}(\bm y;\bm c).
\end{align}
Expanding both sides as series of $\bm x$ and comparing the coefficiencies, we obtain
\begin{align}
  \prod_k y_k^{n_k} = \sum_{2\bm m\leq\bm n}\prod_k\frac{n_k!(c_k/2)^{m_k}}{m_k!(n_k - 2m_k)!}H_{\bm n-2\bm m}(\bm y;\bm c).
\end{align}
Using \Eq{app:hermite-odo}, we know that 
\begin{align}\label{B7}
  {\rm tr}_{\B}\bigg[ \prod_k\mathscr f_k^{\gtrless n_k}\rho_{\T}(t) \bigg]\! =\!\! \sum_{2\bm m\leq\bm n}\prod_k\frac{n_k!(\eta^{\gtrless}_k/2)^{m_k}}{m_k!(n_k - 2m_k)!}\rho_{\bm n-2\bm m}(t).
\end{align}

Consider the left action. Multiplying \Eq{B7} both sides with 
\begin{align*}
  \prod_k\frac{x_k^{n_k}}{n_k!}
\end{align*}
and summing over all $\bm n$,
we consequently obtain 
\begin{align}\label{B8}
  {\rm tr}_{\B}[e^{\sum_kx_k\mathscr f_k^>}\rho_{\T}(t)] = e^{\sum_k\eta_kx_k^2/2}\sum_{\bm n}\prod_k\frac{x_k^{n_k}}{n_k!}\rho_{\bm n}(t).
\end{align}
The right-hind-side is calculated via
\begin{align}
  {\rm r.h.s.} &= \sum_{\bm n} \prod_k\frac{x_k^{n_k}}{n_k!}\sum_{2\bm m\leq\bm n}\prod_k\frac{n_k!(\eta^{\gtrless}_k/2)^{m_k}}{m_k!(n_k - 2m_k)!}\rho_{\bm n-2\bm m}(t) \nl
  &= \sum_{\bm m,\bm l}\prod_k\frac{(\eta_kx_k^2/2)^{m_k}}{m_k!}\frac{x_k^{l_k}}{l_k!}\rho_{\bm l}(t) \nl
  &= e^{\sum_k\eta_kx_k^2/2}\sum_{\bm n}\prod_k\frac{x_k^{n_k}}{n_k!}\rho_{\bm n}(t).
\end{align}
Let all $x_k = x$ in \Eq{B8} and utilize
\begin{align}
  e^{x\sum_k\mathscr f_k^>}\rho_{\T}(t) = e^{x\hat F^>}\rho_{\T}(t), \quad \sum_k\eta_k = \la\hat F^2\ra_{\B},
\end{align}
and we thus have 
\begin{equation}\label{app:force-reconstruction}
 {\rm tr}_{\B}[\hat F^p\rho_{\T}(t)]
 =\sum_{m=0}^{\lfloor p/2\rfloor}
 \frac{p!\la\hat F_{\B}^2\ra_{\B}^m}{2^m m!(p-2m)!}\hat M_{p-2m}(t)
\end{equation}
with 
\begin{align}
  \hat M_{n}(t) \equiv {\rm tr}_{\B}[\mathcal N(\hat F^n)\rho_{\T}(t)] = \sideset{}{'}{\sum}_{\bm n} \frac{n!}{\prod_kn_k!}\rho_{\bf n}(t).
\end{align}
Here, the prime summation is under the constraint $\sum_kn_k = n$. 

In particular, for a normalized total state,
\begin{subequations}\label{app:low-force-moments}
  \begin{align}
 \la\hat F\ra_t
 &=\sum_k{\rm tr}_{\tS}\rho_{\bm 0_k^+}(t),\\
 \la\hat F^2\ra_t
 &=\la\hat F^2\ra_{\B}+\sum_{ij}{\rm tr}_{\tS}\rho_{\bm 0_{ij}^{++}}(t).
\end{align}
\end{subequations}
Subtracting $\la\hat F\ra_t^2$ from the second line gives the evolving variance. More generally, for any system observable $\hat A$,
\begin{equation}\label{app:hybrid-observable}
 \la\hat A\hat F^p\ra_t
 =\sum_{m=0}^{\lfloor p/2\rfloor}
 \frac{p!\la\hat F^2\ra_{\B}^m}{2^m m!(p-2m)!}
 {\rm tr}_{\tS}[\hat A\hat M_{p-2m}(t)].
\end{equation}
The force characteristic function follows by putting $x=i\lambda$ in \Eq{B8} and taking the trace over the system,
\begin{equation}\label{app:force-characteristic}
 \chi_F(\lambda,t) \equiv {\rm Tr}[e^{i\lambda\hat F}\rho_{\T}(t)]
 =e^{-\frac{\lambda^2}{2}\la\hat F^2\ra_{\B}}
 \sum_{n=0}^\infty\frac{(i\lambda)^n}{n!}
 {\rm tr}_{\tS}\hat M_{n}(t).
\end{equation}
This identity is understood as a moment expansion; reconstructing the full probability distribution requires convergence or a separately controlled resummation. Each finite-order moment in \Eq{app:force-reconstruction}, however, requires only ODOs up to that order, with their dynamical accuracy set by the convergence of the propagated hierarchy.

\section{Coupled Lindblad equation for bath evolution}\label{app:langevin}

Equation~\eqref{dissipaton-langevin} describes a finite set of collective bath modes driven by a common vacuum input. Here we derive the corresponding Lindblad equation for these modes after tracing out the input field $\hat\xi(t)$. The construction uses the quantum input--output correspondence between Langevin and master equations \cite{Gar853761,Com17784}. It is a Markovian representation of the collective modes and their residual white-noise environment, not a dissipative replacement of the unitary evolution of the complete microscopic bath. We assume a finite, linearly independent set of dissipaton kernels, so that $\mathbf M$ is positive definite, and use the vacuum input preparation specified in Sec.~\ref{sec:continuous}.

\subsection{Canonical modes}

Denote $\hat{\bm f}=(\hat f_1,\ldots,\hat f_K)^T$, $\bm r=(r_1,\ldots,r_K)^T$, and $\bm\Gamma=\operatorname{diag}(\gamma_1,\ldots,\gamma_K)$. Since $[\hat\xi(t),\hat\xi^\dagger(t')]=2\delta(t-t')$, define the normalized input increment $\ud \hat B_t=\int_t^{t+\ud t}\ud\tau\,\hat \xi(\tau) / \sqrt 2$ for $\ud t>0$. 
The increment commutes with the mode operators adapted up to time $t$, i.e., $[\ud \hat B_t, \hat{\bm f}_{\B}(t)]=0$. Equation~\eqref{dissipaton-langevin} consequently becomes
\begin{equation}\label{app:vector-langevin}
 \ud\hat{\bm f}_{\B}(t)=-\bm\Gamma\hat{\bm f}_{\B}(t)\ud t
 +\sqrt2\bm r\,\ud \hat B_t.
\end{equation}
Moreover, the matrix $\mathbf M$ satisfies
\begin{equation}\label{app:gram-lyapunov}
 \bm \Gamma \mathbf M+\mathbf M\bm \Gamma^\dagger=2\bm r\bm r^\dagger,
\end{equation}
as follows directly from $(\gamma_i+\gamma_j^*)M_{ij}=2r_i r_j^*$. This is also the condition that \Eq{app:vector-langevin} preserve $[\hat f_i(t),\hat f_j^\dagger(t)]=M_{ij}$.

To proceed, we pass to canonical annihilation operators,
\begin{equation}\label{app:canonical-modes}
 \hat{\bm \alpha}\equiv\mathbf M^{-1/2}\hat{\bm f}
\end{equation}
with $[\hat \alpha_i,\hat \alpha_j^\dagger]=\delta_{ij}$. 
Here $\mathbf M^{1/2}$ is the positive Hermitian square root. In this basis, the generalized Langevin equation becomes
\begin{align}\label{app:canonical-langevin}
  \ud\hat{\bm \alpha}_{\B}(t)=-\tilde{\bm\Gamma}\hat{\bm \alpha}_{\B}(t)\ud t+\sqrt2\tilde{\bm r}\,\ud \hat B_t,
\end{align}
where we define
\begin{align}
 \tilde{\bm\Gamma}&=\mathbf M^{-1/2}\bm\Gamma \mathbf M^{1/2},\quad
 \tilde{\bm r}=\mathbf M^{-1/2}\bm r.
\end{align}
Multiplying \Eq{app:gram-lyapunov} by $\mathbf M^{-1/2}$ on both sides gives $\tilde{\bm\Gamma}+\tilde{\bm\Gamma}^\dagger=2\tilde{\bm r}\tilde{\bm r}^\dagger$. Thus the Hermitian and anti-Hermitian parts of $\tilde{\bm\Gamma}$ specify the damping and Hamiltonian matrices, respectively:
\begin{equation}\label{app:canonical-drift-split}
 \tilde{\bm\Gamma}=\tilde{\bm r}\tilde{\bm r}^\dagger + i\bm\Omega,\quad
 \bm\Omega\equiv \frac{\tilde{\bm\Gamma}-\tilde{\bm\Gamma}^\dagger}{2i}=\bm\Omega^\dagger.
\end{equation}

\subsection{Hamiltonian, jump operator, and master equation}

The time dependence of $\hat{\bm \alpha}_{\B}(t)$ can be interpreted from two aspects. On the one hand, it is the Heisenberg evolution of the canonical modes under a unitary operator $U_{\B}(t) = e^{-iH_{\B}t}$. On the other hand, it is the solution of a quantum stochastic differential equation, \Eq{app:canonical-langevin}. 

For the quantum stochastic process, we can define the corresponding evolution operator\label{app:unitary-qsde}
\begin{align}
  \ud G_{\B}(t) \equiv \left[\hat L\ud \hat B_t^\dagger-\hat L^\dagger\ud \hat B_t
 -\left(i\hat H_{\rm aux}+\frac12\hat L^\dagger\hat L\right)\ud t\right]G_{\B}(t)
\end{align}
with the quadratic Hamiltonian and the jump operator being
\begin{align}\label{app:canonical-HL}
 \hat H_{\rm aux}\equiv\hat{\bm \alpha}^\dagger\bm\Omega\hat{\bm \alpha},\quad
 \hat L\equiv-\sqrt 2\tilde{\bm r}^\dagger\hat{\bm \alpha},
\end{align}
respectively. After imposing the quantum It\^o rules, 
\begin{equation}\label{app:vacuum-ito}
 \ud \hat B_t\ud \hat B_t^\dagger=\ud t,\quad 
 \ud \hat B_t^\dagger\ud \hat B_t=(\ud \hat B_t)^2=0,
\end{equation}
it is easy to verify that $\ud[G_{\B}^\dagger(t)G_{\B}(t)] = 0$. For any mode observable $\hat A$, its time evolution is given by
\begin{align}
  \hat A_{\B}(t) \equiv  G_{\B}^\dagger(t)\hat A G_{\B}(t).
\end{align}
Substituting \Eq{app:unitary-qsde} and using the quantum It\^o rule \Eq{app:vacuum-ito}, we obtain 
\begin{align}\label{app:observable-qsde}
  \ud\hat A_{\B}(t)={}&G^\dagger_{\B}(t)\mathcal D^\dagger\hat AG_{\B}(t)\ud t
 +[\hat L^\dagger_{\B}(t),\hat A_{\B}(t)]\ud \hat B_t
  \nl
  &+\ud \hat B_t^\dagger[\hat A_{\B}(t),\hat L_{\B}(t)],
\end{align}
with
\begin{align}
  \mathcal D^\dagger(\hat A)\equiv {}&i[\hat H_{\rm aux},\hat A]
 +\hat L^\dagger\hat A\hat L
 -\frac12\{\hat L^\dagger\hat L,\hat A\}.
\end{align}
In particular, $[\hat L^\dagger,\hat{\bm \alpha}]=\sqrt2\tilde{\bm r}$ and $[\hat{\bm \alpha},\hat L]=0$, while
\begin{equation}\label{app:canonical-adjoint-drift}
 \mathcal D^\dagger(\hat{\bm \alpha})
 =-\left(i\bm\Omega+\tilde{\bm r}\tilde{\bm r}^\dagger\right)\hat{\bm \alpha}
 =-\tilde{\bm\Gamma}\hat{\bm \alpha}.
\end{equation}
Thus this stochastic evolution reproduces both the drift and the input noise of \Eq{app:canonical-langevin}.

Let $\varrho_{\B}(t)$ be the density operator of the $K$ canonical modes after tracing out the input field. Averaging \Eq{app:observable-qsde} over the vacuum removes the stochastic increments. Equivalently, expanding $U_{\B}(t)\varrho_{\B}(0)U_{\B}^\dagger(t)$ with the It\^o product rule produces the recycling term $\hat L\varrho_{\B}\hat L^\dagger\ud t$. The resulting master equation is
\begin{equation}\label{app:auxiliary-lindblad}
 \dot\varrho_{\B}=-i[\hat H_{\rm aux},\varrho_{\B}]
 +\hat L\varrho_{\B}\hat L^\dagger
 -\frac12\{\hat L^\dagger\hat L,\varrho_{\B}\}.
\end{equation}
This has the standard Lindblad form \cite{Lin76119}. Its vacuum input dilation specifies a completely positive, trace-preserving evolution on the mode density operators. 

\bibliography{./ref.bib}

\end{document}